\documentclass[journal=jpca,manuscript=article, layout=twocolumn]{achemso}

\usepackage{chemformula} % Formula subscripts using \ch{}
\usepackage[T1]{fontenc} % Use modern font encodings

\usepackage{graphicx}% Include figure files
\usepackage{dcolumn}% Align table columns on decimal point
\usepackage{bm}% bold math
\usepackage[mathlines]{lineno}% Enable numbering of text and display math
\usepackage{pgfplots}
\usepackage{amsmath,amssymb,amsfonts}
\usepackage{mathtools}
\usepackage{algorithmic}
\usepackage{subcaption}
\usepackage{gensymb}
\usepackage{comment}
\usepackage[utf8]{inputenc}
\usepackage[T1]{fontenc}
\usepackage{mathptmx}
\usepackage{etoolbox}
\usepackage{mathtools}
\usepackage{amsmath,amssymb,amsfonts}
\usepackage{cuted}
\usepackage{lipsum}
\usepackage{booktabs}
\usepackage{xcolor}

\usepackage[braket, qm]{qcircuit}
\usepackage{textcomp}
\usepackage{xcolor}
\usepackage{braket}
\usepackage{physics}
\usepackage{chemfig}
\usepackage{chemformula}

\usepackage{soul}

\usepackage{bibunits}
\usepackage{xr}
\usepackage{cleveref}
\usepackage{caption}
\graphicspath{{figures/}}
\usepackage{epstopdf}%This line makes .eps figures into .pdf - please comment out if not required.


\newcommand{\romnum}[1]{\lowercase\expandafter{\romannumeral #1\relax}}
\newcommand{\angstrom}{\text{\normalfont\AA}}

\author{Nirmal Mammavalappil Rajan}
 \affiliation{TCS Research, Tata Consultancy Services Limited, Bengaluru, India-560066}%Lines break automatically or can be forced with \\
\author{Ankit Khandelwal}%
 \affiliation{TCS Research, Tata Consultancy Services Limited, Bengaluru, India-560066}
\author{Manoj Nambiar}
 \affiliation{TCS Research, Tata Consultancy Services Limited, Mumbai, India-400076}
\author{Sharma S. R. K. C. Yamijala}
 \affiliation{Department of Chemistry, Indian Institute of Technology Madras, Chennai, India-600036}
 \alsoaffiliation{Centre for Quantum Information, Communication, and Computing, Indian Institute of Technology Madras, Chennai, India-600036}
 \email{yamijala@iitm.ac.in}

\title[Parallelized Givens Ansatz]{Parallelized Givens Ansatz for Molecular ground-states: Bridging Accuracy and Efficiency on NISQ Platforms}
\abbreviations{VQE, UCCSD, PGSD, AS}
\SectionNumbersOn

\begin{document}

\begin{abstract}
In recent years, the Variational Quantum Eigensolver (VQE) has emerged as one of the most popular algorithms for solving the electronic structure problem on near-term quantum computers. The utility of VQE is often hindered by the limitations of current quantum hardware, including short qubit coherence times and low gate fidelities. These limitations become particularly pronounced when VQE is used along with deep quantum circuits, such as those required by the "Unitary Coupled Cluster Singles and Doubles" (UCCSD) ansatz, often resulting in significant errors. To address these issues, we propose a low-depth ansatz based on \emph{parallelized} Givens rotations, which can recover substantial correlation energy while drastically reducing circuit depth and two-qubit gate counts for an arbitrary active space (AS). Also, considering the current hardware architectures with low qubit counts, we introduce a systematic way to select molecular orbitals to define active spaces (ASs) that retain significant electron correlation. We validate our approach by computing bond dissociation profiles of water and strongly correlated systems, such as molecular nitrogen and oxygen, across various ASs. Noiseless simulations using the new ansatz yield ground-state energies comparable to those from the UCCSD ansatz while reducing circuit depth by 50–70\%. Moreover, in noisy simulations, our approach achieves energy error rates an order of magnitude lower than that of UCCSD. Considering the efficiency and practical usage of our ansatz, we hope that it becomes a potential choice for performing quantum chemistry calculations on near-term quantum devices.
\end{abstract}

%\maketitle

\section{\label{sec:Introduction}Introduction}
With the recent demonstration of potential advantage and utility of near-term quantum computers,\cite{kim2023evidence, arute2019quantum, zhong2020quantum, wu2021strong} the field of quantum computation has attracted immense research interest across various scientific domains. Since its original proposition, simulation of chemical systems remains one of the most promising applications of this emerging computing platform.\cite{mcardle2020quantum, cao2019quantum, motta2022emerging, bauer2020quantum} A variety of quantum algorithms have been developed over the past few decades to study many-electron systems using quantum computers. Some of the prominent examples include the Quantum Phase Estimation (QPE) \cite{abrams1999quantum, aspuru2005simulated}, Quantum Subspace Expansion (QSE) \cite{mcclean2017hybrid, colless2018computation}, Variational Quantum Eigensolver (VQE) \cite{peruzzo2014variational, cerezo2021variational}, and Imaginary Time Evolution \cite{motta2020determining}. Among these, the VQE algorithm has emerged as the most widely used approach for estimating molecular properties on present-day quantum computers. 

In the VQE algorithm, the ground-state wavefunction of a molecule is represented using a quantum circuit (often referred to as an ansatz) comprising of several quantum gates with tunable parameters. These parameters are iteratively optimized using a classical algorithm such that the optimized quantum circuit provides the best estimate of the molecular ground-state energy achievable with the chosen ansatz. Therefore, on an ideal quantum computer, the accuracy of the VQE algorithm is mainly governed by the choice of the ansatz.\cite{mcclean2016theory}

In general, there are two groups of ansatzes, namely, chemically inspired, and hardware-efficient ansatzes, where the latter ansatz usually consists of alternating layers of parameterized single-qubit rotations and two-qubit entangling gates that are more tailored to the underlying quantum hardware \cite{kandala2017hardware}. In contrast, a chemically inspired ansatz typically relies on the unitary parametrization of the existing quantum chemistry methods. For example, the family of Unitary Coupled Cluster (UCC) ansatzes \cite{romero2018strategies, anand2022quantum} are inspired from the classical coupled cluster approaches. In fact, the first experimental demonstration of the VQE algorithm \cite{peruzzo2014variational} utilized the UCC ansatz restricted to singles and doubles, i.e., UCCSD. Mathematically, the UCCSD ansatz can be written as
\begin{equation}
    \ket{\Psi_{UCCSD}} = e^{T-T^{\dagger}}\ket{\Psi_{ref}}
    \label{eqn:Psi UCCSD}
\end{equation}
where $T=\sum_{i,\alpha} t_i^\alpha a_\alpha^\dagger a_i + \sum_{i,j,\alpha,\beta} t_{ij}^{\alpha\beta} a_\alpha^\dagger a_\beta^\dagger a_i a_j$ is the second-order cluster operator that accounts for the electronic excitations from occupied ($i,j,..$) to unoccupied orbitals ($\alpha,\beta,..$) in a reference wavefunction $\ket{\Psi_{ref}}$. Often, the Hartree-Fock  (HF) wavefunction is chosen as the reference state, i.e., $\ket{\Psi_{ref}} = \ket{\Psi_{HF}}$. Since the UCCSD ansatz is derived from the traditional coupled cluster method, it possesses a few essential features, such as size consistency and size extensivity. In addition, due to the unitary parametrization, the UCCSD ansatz also obeys the variational principle. 

While implementing the exponentiated cluster operator on a quantum computer, generally, the Trotter approximation \cite{nielsen2010quantum} is used to account for the non-commutativity of the constituent excitation operators. Interestingly, even with a single Trotter step, VQE simulations with UCCSD ansatz often provides highly accurate energies on ideal quantum simulators, even for systems with strong static correlation.\cite{barkoutsos2018quantum, sokolov2020quantum, rossmannek2021quantum} However, unfortunately, the quantum computational cost associated with the implementation of UCCSD ansatz scales polynomially with system size. For instance, the circuit depth of a Trotterized UCCSD ansatz scales as $O(\tau N_{occ}^2 N_{vir}^2 N)$ \cite{motta2023bridging}, and the number of two-qubit gates as $O(N^4\tau)$, where $\tau$ is the number of Trotter steps used to approximate the exponential cluster operator, $N_{occ}$, $N_{vir}$, and $N (=N_{occ}+N_{vir})$ are the occupied, virtual, and total number of spin-orbitals of the $\ket{\Psi_{ref}}$, respectively. A more general expression for the scaling of UCCSD circuit depth is given by $O((N-m)^2m\tau)$, where $m$ is the number of electrons, and that of other commonly employed ansatzes within VQE are discussed in the reference.\citenum{tilly2022variational}  

Consequently, reliable execution of quantum simulations using the UCCSD ansatz is impractical on the noisy intermediate scale quantum (NISQ) devices due to numerous hardware limitations. These include short coherence times ($\approx 100$ \textmu s for superconducting platforms), high readout and two-qubit gate errors, and limited qubit connectivity.\cite{ibmquantumplatform} Recent experiments on superconducting quantum hardware have reported energy errors of approximately 1 Hartree while predicting the ground-state energies of H\textsubscript{2} and LiH molecules using two- and four-qubit UCCSD circuits.\cite{ghosh2023deep}

To address these limitations, it is crucial to develop ansatzes that minimize the consumption of quantum resources while providing highly accurate results. Towards this goal, various research groups have proposed different design strategies for ansatz construction. These include (i) designing ansatzes that are tailored to the chemical system, where operators that can significantly recover the correlation energy are systematically incorporated into the quantum circuit. Examples of this class of ansatzes are the Qubit Coupled Cluster (QCC) approach \cite{ryabinkin2018qubit, ryabinkin2020iterative}, and ADAPT-VQE \cite{grimsley2019adaptive, tang2021qubit, yordanov2021qubit}; (ii) enforcing various symmetry elements of the system such as particle number, spin-symmetry, and molecular symmetry, to restrict the VQE search space from $2^{N}$-dimensional Hilbert space to a low-dimensional subspace. Examples of this class of ansatzes include various symmetry-preserving ansatzes \cite{gard2020efficient, barkoutsos2018quantum, roth2017analysis, cao2022progress}; (iii) introducing ancilla qubits (as considered in quantum neural networks (QNN)) to reduce the circuit depth \cite{zeng2023quantum}; (iv) unitary parametrization of the cluster Jastrow (uCJ) method and its variants \cite{matsuzawa2020jastrow, motta2023bridging}; (v) decomposition approaches, where many smaller quantum circuits are used to model the fragments of a larger system, and the results from smaller circuits are later combined (often, using a classical computer) to obtain the physical properties of the larger system of our interest. Examples include entanglement forging based on Schmidt decomposition \cite{eddins2022doubling, motta2023quantum}, and circuit cutting-knitting techniques \cite{peng2020simulating, lowe2023fast}.

In this work, we propose a novel method for designing highly efficient ansatzes for the VQE algorithm. Our approach is based on applying Givens rotation gates in a parallel manner for larger ASs. We show that the designed quantum circuits systematically reduce the circuit depth and two-qubit gate count compared to the UCCSD ansatz, while retaining ground-state energies closer to the chemical accuracy. We refer to these ansatzes as parallelized Givens singles and doubles (PGSD). Furthermore, these PGSD circuits are more noise resilient than the UCCSD circuits, i.e., the PGSD ansatz can predict better quality energy profiles even in the presence of device noise. In addition to the PGSD ansatz, considering the current hardware architectures with low qubit counts, we present a simple yet very useful method for selecting ASs that can recover the majority of the correlation energy for a given molecular conformation. We experimentally demonstrate the effectiveness of these methods by simulating the bond dissociation profiles of a few strongly correlated systems.

The remainder of this paper is organized as follows: In Sec.~\ref{subsec:Givens Ansatz}, we describe the procedure for constructing quantum circuits as a sequence of Givens rotation-based gates. We start with a simple example and later generalize it to any arbitrary AS. For larger ASs, we provide a procedure to apply these gates in parallel for executing these circuits efficiently on NISQ devices.  In Sec.~\ref{subsec:ASS-CCSD}, we outline a methodology to choose an appropriate AS that can retain a significant portion of electron correlation. In Sec.~\ref{sec: Results}, we present and analyze the results obtained by applying both these methods for simulating the potential energy surfaces (PES) of small molecules, such as water, molecular nitrogen, and molecular oxygen, under noiseless conditions. Furthermore, to assess the practical relevance of the PGSD ansatz on current quantum hardware, we compared its performance against the UCCSD ansatz in the presence of device noise. We conclude the article with Sec.~\ref{sec: Conclusion}.

\section{\label{sec:Methodology} Methodology} 
\subsection{\label{subsec:Givens Ansatz} Preparing the Ground-state using Givens rotation-based gates}
To accurately determine the ground-state wavefunction of a molecular system, we use the VQE algorithm, which is based on the Rayleigh-Ritz variational principle.\cite{peruzzo2014variational, kandala2017hardware} We use the Jordan-Wigner (JW) mapping \cite{seeley2012bravyi} to represent the electronic Hamiltonian and wavefunction in the qubit basis, \textit{without} tapering qubits based on $\textit{Z\textsubscript{2}}$ symmetries.\cite{bravyi2017tapering} Under the JW mapping, each molecular spin orbital corresponds to a qubit, where the single-qubit states $\ket{0}$ and $\ket{1}$ describe the occupation of the spin orbital. Therefore, $2M$ qubits are needed to describe the electronic wavefunction of a system with $M$ molecular orbitals. For clarity, we adopt the convention that the first $M$ qubits ($q_{0}, q_{1},..q_{M-1}$) represent the spin-up orbitals and the remaining $M$ qubits ($q_{M}, q_{M+1},..q_{2M-1}$) represent the spin-down orbitals. Our aim is to design an ansatz for the VQE algorithm that can be efficiently executed on NISQ devices. We begin by demonstrating the procedure to construct an ansatz using the Givens rotation-based gates for a simple yet non-trivial example: a fermionic system with two molecular orbitals (MOs) and two electrons. We then extend the approach to any arbitrary AS comprising of $M$ molecular orbitals and $2N$ electrons.

\subsubsection{An example system with 2 MOs and 2 electrons}
\begin{figure}[h]
    %\centering
    \includegraphics[width=0.9\columnwidth]{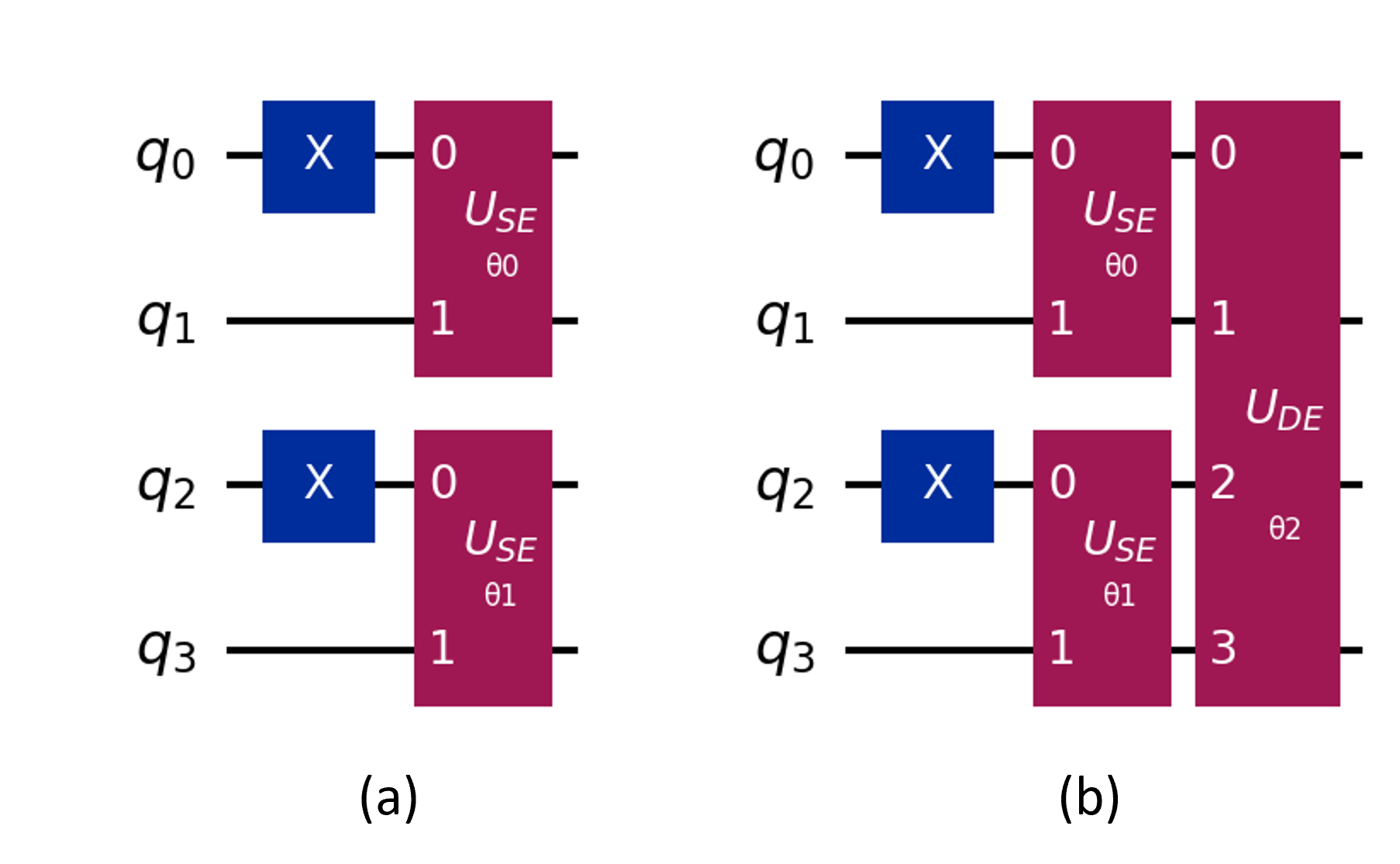}
    \caption{Four-qubit Givens circuits representing the ground-state of a fermionic system with two MOs and two electrons. A circuit with (a) Pauli-X and single excitation gates ($U_{SE}$) representing the wavefunction defined in Eq.~\ref{eq: Psi_S ansatz}. (b) Pauli-X, single ($U_{SE}$) and double ($U_{DE}$) excitation gates representing the wavefunction defined in Eq.~\ref{eq: Psi_SD}. }
    \label{fig1: 4-qubit Givens circuits}    
\end{figure}

A quantum circuit that prepares the ground-state of our example system (with two MOs and two electrons) using the Givens rotation-based gates is shown in Fig.~\ref{fig1: 4-qubit Givens circuits}. Here, since the system has two MOs, we considered a 4-qubit circuit. Also, as per the orbital encoding scheme mentioned above, the qubits $q_{0}$ and $q_{1}$ ($q_{2}$ and $q_{3}$) represent the spin-up (down) orbitals, with $q_{0}$ ($q_{2}$) being lower in energy than $q_{1}$ ($q_{3}$). We begin the state preparation by applying Pauli X gates on qubits $q_0$ and $q_2$ to initialize the 4-qubit state in the HF singlet state, i.e., $\Psi_{HF} \equiv I_3X_2I_1X_0\ket{0000} =\ket{0101}$. The electronic excitations contributing to the correlation energy include two single excitations, represented by the states |0110⟩ and |1001⟩, and a double excitation represented by the state |1010⟩. To simulate these excitations on the qubits, we utilize the Givens rotation matrices.\cite{anselmetti2021local, arrazola2022universal} 

%\textcolor{red}{\textit{Describing the operation and the particle-conserving nature of the single-excitation gate, which is a building block in our Givens ansatz.}} 
For example, to create the single excitations |0110⟩ and |1001⟩, one can consider a parameterized two-qubit gate $U_{SE}(\theta)$ described by the unitary matrix
\begin{equation}
    U_{SE}(\theta) = \begin{pmatrix} 
    1 & 0 & 0 & 0\\
    0 & \cos({\theta}) & \sin({\theta}) & 0\\
    0 & -\sin({\theta}) & \cos({\theta}) & 0\\
    0 & 0 & 0 & 1\\
    \end{pmatrix}
    \label{eqn: U_se matrix}
\end{equation}
in the computational basis $\set{\ket{00}, \ket{01}, \ket{10},\ket{11}}$. The action of $U_{SE}(\theta)$ can be understood by applying it on a two-qubit basis state $\ket{01}$, where the first spin orbital is unoccupied and the second one is occupied. As shown below, applying $U_{SE}(\theta)$ on $\ket{01}$ returns a normalized superposition of the original state, $\ket{01}$, and the singly-excited state, $\ket{10}$.
\begin{equation}
    \centering
    U_{SE}(\theta)\ket{01} = \cos(\theta)\ket{01}-\sin(\theta)\ket{10},
    \label{eq: action of SE}
\end{equation}
Here, the real parameter $\theta$ controls the amplitude of the excited states. A similar action can be observed when $U_{SE}(\theta)$ is applied to the basis state $\ket{10}$. However, when this gate is applied on either $\ket{00}$ or $\ket{11}$ (corresponding to the states with zero and two electrons, respectively), it will not return a superposition state. In other words, $U_{SE}(\theta)$ is an excitation-preserving gate (i.e., it preserves the particle number), which is one of the important characteristics of Givens matrices. Another way of interpreting $U_{SE}(\theta)$ is that it performs a rotation in the two-dimensional subspace spanned by the basis states $\ket{01}$ and $\ket{10}$ of the four-dimensional two-qubit Hilbert space. An implementation of $U_{SE}(\theta)$ as a sequence of standard single-qubit and two-qubit gates used for our simulations is shown in Fig.~\ref{supp-figS1: SE gate} and is taken from the reference \citenum{anselmetti2021local}.

%\textcolor{red}{\textit{Describing the output of the 4-qubit Givens ansatz with SE gates only and why we need to introduce a DE gate.}}
The trial state prepared by the circuit in Fig.~\ref{fig1: 4-qubit Givens circuits}(a) after the application of the two single excitation (SE) gates can be written as 
\begin{equation}
    \centering
    \begin{split}
        \ket{\Psi_{S}(\mathbf{\theta})}  \equiv& U_{SE}(\theta_1) U_{SE}(\theta_0) \ket{0101} \\ 
         = &c_0c_1\ket{0101}-c_0s_1\ket{1001} \\
        &   -s_0c_1\ket{0110}+s_0s_1\ket{1010},
    \end{split}
    \label{eq: Psi_S ansatz}
\end{equation}
where $c_i=\cos(\theta_i)$ and $s_i=\sin(\theta_i)$. Although the application of the SE gates leads to the formation of both single ($\ket{0110}$ and $\ket{1001}$) and doubly excited states ($\ket{1010}$), only some of these excited states are accessible through the circuit in Fig.~\ref{fig1: 4-qubit Givens circuits}(a). To understand this issue, let us consider the H\textsubscript{2} molecule as an example. In the minimal basis, its exact ground-state is a linear combination of the HF state and the doubly excited state. Therefore, to achieve the exact ground-state of H\textsubscript{2}, the coefficients of the singly excited states must become zero. However, to make the coefficient of $\ket{1001} (\ket{0110})$ as zero, either $c_0$ or $s_1$ ($s_0$ or $s_1$) must be zero, but this also makes the coefficients of either the HF or the doubly excited state zero. In other words, the circuit with only singly excited states is not sufficient to obtain the true ground-state of the H\textsubscript{2} molecule. 

The above discussion motivates us to introduce a parameterized four-qubit double-excitation gate that can linearly combine four-qubit basis states $\ket{0101}$ and $\ket{1010}$, while leaving all other basis states unaffected. This Givens gate will exchange two electrons between two occupied and two virtual spin orbitals. One form of such a unitary can be described by the relations 
\begin{equation}
    \centering
    \begin{split}
        U_{DE}(\theta)\ket{0101} = \cos(\theta)\ket{0101}+\sin(\theta)\ket{1010} \\
        U_{DE}(\theta)\ket{1010} = \cos(\theta)\ket{1010}-\sin(\theta)\ket{0101}
    \end{split}
    \label{eq: action of DE1}
\end{equation}
which, like the unitary $U_{SE}(\theta)$, is particle-conserving and only performs a rotation in the two-dimensional subspace spanned by states $\ket{0101}$ and $\ket{1010}$ of the sixteen-dimensional Hilbert space. As noted in the literature \cite{arrazola2022universal}, double excitation gates can be constructed to perform rotations in other two-dimensional subspaces defined by states $\set{\ket{0011}, \ket{1100}}$, and $\set{\ket{0110}, \ket{1001}}$. The decomposition of the unitary $U_{DE}(\theta)$ in terms of standard single- and two-qubit gates is given in Fig.~\ref{supp-figS2: DE gate} and it is a modified version of the circuit presented in reference \citenum{arrazola2022universal} to account for the chosen ordering of qubits. 
% Various other implementations of the double-excitation gates and their quantum resource estimates are given in Supplementary Info.

%\textcolor{red}{\textit{Describing the output of the 4-qubit Givens ansatz with two SE gates and one DE gate and some observations.}}
%Note that we have carefully applied the double excitation gate $U_{DE}$ such that it exchanges electrons between the initially excited qubits $q_0$ and $q_2$ and unexcited qubits $q_1$ and $q_3$.
The four-qubit Givens ansatz comprising two single and one double excitation gates is shown in Fig.~\ref{fig1: 4-qubit Givens circuits}(b). The trial state prepared by this quantum circuit can be expressed as  
\begin{align}
    \ket{\Psi_{SD}(\mathbf{\theta})} \equiv& U_{DE}(\theta_2) U_{SE}(\theta_1) U_{SE}(\theta_0) \ket{0101} \notag\\
    =& (c_0c_1c_2+s_0s_1s_2)\ket{0101}\notag\\
    &- (c_0c_1s_2-s_0s_1c_2)\ket{1010} \notag \\ 
    &- c_0s_1\ket{1001}-s_0c_1\ket{0110}.
    \label{eq: Psi_SD}
\end{align}
Clearly, in Eq.~\ref{eq: Psi_SD}, the amplitudes of the first two terms can now be controlled by tuning the parameter $\theta_2$, and hence, justifies the necessity to employ the double excitation gates while preparing the quantum circuits of molecules, as observed previously by Anselmetti \emph{et al.} \cite{arrazola2022universal} . Our simulations with the circuit in Fig.~\ref{fig1: 4-qubit Givens circuits}(b) indicate that this circuit can prepare the ground-state of H\textsubscript{2} molecule in a minimal basis for a large range of interatomic distances with chemical accuracy. 

\begin{figure*}[!htb]
    \centering
    \includegraphics[width=\linewidth]{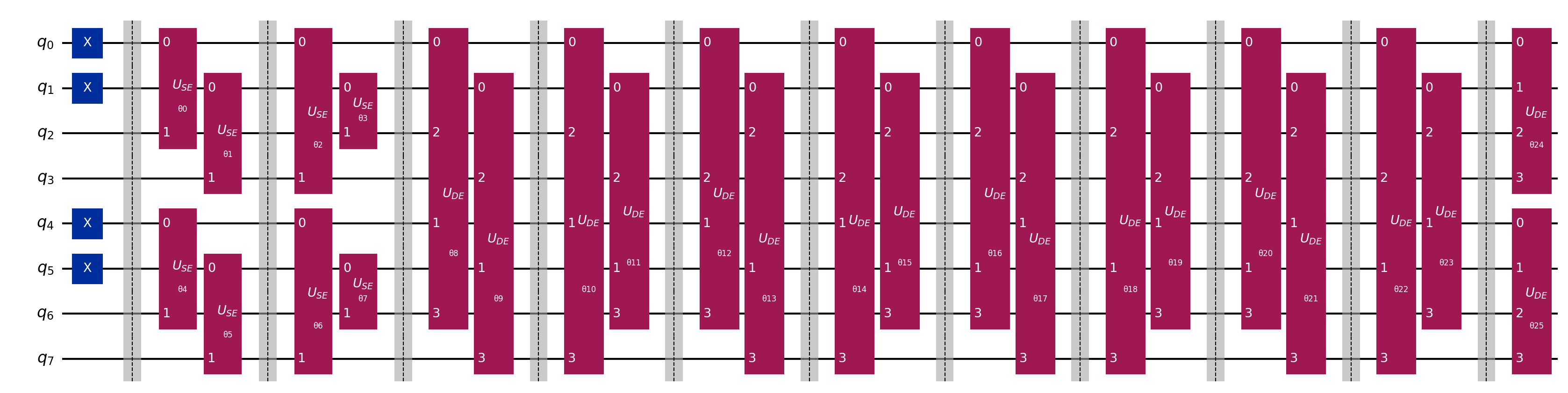}
    \caption{An 8-qubit PGSD circuit designed for an AS(4e, 4o) in which commuting single and double excitation gates are applied in one layer (marked by two dashed barriers). At this level of representation, the PGSD ansatz has a depth of 12.}
    \label{fig:CAS(4e, 4o) PGSD}
\end{figure*}

\subsubsection{Preparing the Givens ansatz for an arbitrary active space}
% {- counting the number of single and double excitations.}
After successfully demonstrating the construction of a four-qubit Givens ansatz for a system with two molecular orbitals and two particles, we extend our approach to a general active space comprising of $2N$ electrons and $M$ molecular orbitals, denoted by AS($2N$e, $M$o). Without loss of generality, we fix the number of $\alpha$- and $\beta$-spin electrons to be equal i.e., $N_{\alpha}=N_{\beta}=N$. In section \ref{supp-sec: Number of possible excitations}, we present the equations for finding the number of possible electronic excitations of each order (say, single, double, etc.) in a given active space AS($2N$e, $M$o). 

Once we determine the number of determinants corresponding to each kind of electronic excitation (say, $S$ number of single excitations, $D$ number of double excitations, etc.), the next step is to apply quantum gates that can generate these excitations on the qubit states. Our general protocol to build circuits using Givens rotation gates for an arbitrary active space, AS($2N$e, $M$o), can be outlined as follows:
\begin{enumerate}
    \item Initialize the $2M$-qubit HF state by applying $2N$ Pauli X gates, where $N$ of them will act on the first $N$ qubits encoding the \emph{spin-up} orbitals ($q_0, q_1,...,q_{N-1}$), and the remaining will act on the first $N$ qubits encoding the \emph{spin-down} orbitals ($q_{M},q_{M+1},...,q_{M+N-1}$).
    
    \item Apply $S$ single excitation gates, each with a tunable parameter $\theta_s$ such that each gate acts on a distinct pair of qubits $(q_i,q_a)$, where $q_i$ is in state $\ket{1}$ and $q_a$ in state $\ket{0}$. Wherever possible the single excitation gates applied in one layer are ensured to commute with each other, i.e., gates $S1$ and $S2$, acting on qubits $(q_{i}, q_{a})$ and $(q_{j}, q_{b})$, respectively, are applied in the same layer only when $q_{i} \neq q_{j}$ and $q_{a} \neq q_{b}$. In other words, $S1$ and $S2$ are applied in parallel.
    
    \item Apply $D$ double excitation gates, each with a tunable parameter $\theta_d$ such that each gate acts on a distinct set of four qubits $(q_i,q_a,q_j,q_b)$, where $\ket{q_iq_j}=\ket{11}$ and $\ket{q_aq_b}=\ket{00}$. Once again, wherever possible the double excitation gates applied in one layer are ensured to commute with each other, i.e., gates $D1$ and $D2$, acting on qubits $(q_{i}, q_{a}, q_{j}, q_{b})$ and $(q_{k}, q_{c}, q_{l}, q_{d})$, respectively, are applied in the same layer only when an occupied (unoccupied) qubit in $(q_{i}, q_{a}, q_{j}, q_{b})$ is \emph{not} equal to any of the occupied (unoccupied) qubits in $(q_{k}, q_{c}, q_{l}, q_{d})$. In other words, $q_{i} \neq q_{k}$ or $q_{l}$, $q_{a} \neq q_{c}$ or $q_{d}$, $q_{j} \neq q_{k}$ or $q_{l}$, and $q_{b} \neq q_{c}$ or $q_{d}$.   Thus, the total number of variational parameters in the circuit would be equal to $S+D$.
\end{enumerate}
It is worth emphasizing that applying commuting gates together in one layer has reduced the depth of Givens circuits drastically. Earlier works on UCC and its disentangled form have already shown that the ordering of unitary operators in quantum circuits has a strong effect on the accuracy of the calculations.\cite{evangelista2019exact, izmaylov2020order} Therefore, using the procedure outlined above, we can construct efficient quantum circuits for a system with arbitrary AS. Hereafter, we refer to the circuits that are constructed by applying only Givens single and double excitation gates in a parallel manner as "Parallelized Givens Singles and Doubles" (PGSD). An example 8-qubit PGSD circuit demonstrating the parallel application of Givens excitation gates is shown in Fig.~\ref{fig:CAS(4e, 4o) PGSD} for an AS(4e, 4o). Here, we want to note that, although we did not explicitly apply triple excitation gates in this circuit, the presence of several layers of single and double excitation gates implicitly introduces triple and higher-order excited states (as is commonly observed with the UCCSD circuits). 

In the following, we summarize the prior-art works that utilize particle-conserving unitaries for VQE ansatz design similar to the Givens rotation gates employed in our work. Yordanov \emph{et al.}\cite{yordanov2020efficient} provides both a theoretical description and circuit implementations of fermionic excitations, which are used in UCC ansatzes, and qubit excitations that are widely explored to construct shallow quantum circuits. The qubit excitations are a simplified form of the fermionic excitations in the sense that the exponentials of the product of Pauli-Z gates accounting for fermionic anticommutation relations are omitted leading to the removal of "CNOT-staircases" in the circuit implementation of qubit excitation operators.\cite{magoulas2023cnot} Hence, the gate complexity of the qubit excitation operators is upper-bounded to $O(1)$ while that of fermionic excitations is $O(N)$, lowering the circuit depth and two-qubit gates significantly. The classical optimizer is poised with the task of adjusting the circuit parameters such that the wavefunction prepared by ansatzes employing qubit excitations respects the anti-commutation relations.\cite{xie2022qubit} The unitary matrix representations of these qubit excitations correspond to the Givens rotation matrices. Such qubit excitations are used in the construction of qubit coupled cluster singles and doubles ansatz (QCCSD)\cite{xia2020qubit} (we note here that there exists a different ansatz design protocol with the same name mentioned in Sec. \ref{sec:Introduction}). However, unlike the parallel ordering of the excitation gates in our PGSD ansatz, the QCCSD ansatz follows a sequential arrangement of the qubit excitations applied on the HF state.

Other notable works include that of Xie \emph{et al.} \cite{xie2022qubit} in which the fermionic excitations of the k-UpCCGSD ansatz are replaced by qubit excitations to yield a low-depth ansatz known as k-QUpCCGSD. As observed in this work, the UCC-type ansatzes, such as UCCSD, k-UpCCGSD, are found to be slightly more accurate than the qubit excitation-based ansatzes, such as QUCCSD, k-QUpCCGSD. However, this improvement in accuracy comes at an enormous cost in terms of circuit depth and two-qubit gate count, which are the limiting factors to run quantum computations in the NISQ era. Qubit excitations are also employed in combination with ADAPT-VQE protocols (called QEB-ADAPT-VQE \cite{yordanov2021qubit}) for more efficient problem-tailored ansatz construction methods.

\subsection{\label{subsec:ASS-CCSD}Selection of Active Orbitals}
In the previous section, we have outlined the procedure to construct PGSD circuit for an arbitrary AS. Since the choice of AS hugely dictates the accuracy of the ground-state energy computed using the VQE algorithm (through Hamiltonian), in this section, we will provide a procedure to identify the best AS that can capture large amount of correlation energy. Later we will demonstrate that when an appropriate AS is chosen, the PGSD ansatz can recover large amount of correlation energy even for structures that are near the bond dissociation limit (i.e., far away from the equilibrium geometry). Our approach to select an AS containing significant amount of correlation energy is as follows:
\begin{enumerate}
\item Run a CCSD calculation including all the orbitals in the chosen basis set for the specified geometry
\item Obtain all the single ($t_i^a$) and double ($t_{ij}^{ab}$) excitation amplitudes and the indices $i,j,..$ ($a,b,..$) of the occupied (virtual) orbitals corresponding to each single and double excitation
\item Sort the lists of single and double excitations according to their excitation amplitudes. These sorted lists of single and double excitations, denoted as $S$ and $D$, respectively, help to identify the dominant electronic excitations (ones with larger amplitudes) and can serve as reference datasets for choosing an AS with high correlation energy
\item Consider all possible combinations of ASs with $2N$ electrons and $M$ orbitals that can be formed with all the available orbitals, and for each AS compute the correlation factor $\epsilon$ defined as

\begin{equation}
    \centering
    \epsilon = \sum_{t_i^a, t_{ij}^{ab}} |t_i^a|^2 + |t_{ij}^{ab}|^2 
    \label{eqn: correlation factor}
\end{equation}

where $t_i^a \in S$ and $t_{ij}^{ab} \in D$ are the amplitudes corresponding to all the possible single and double excitations in the chosen AS. The larger the correlation factor of an AS, the higher the amount of electron correlation present in it.
\end{enumerate}

In the following, we discuss the computational bottlenecks associated with this approach and ways to mitigate them. One major complexity of this approach is associated with the requirement of running a full CCSD calculation (i.e., by including all the orbitals for a chosen basis set), which has a scaling of $O(N^6)$, where $N$ is the number of basis functions), and hence, might not be feasible for larger molecules. This can be overcome to a good extent by using less expensive but accurate post-HF methods like classical second-order Moller-Plesset (MP2) or configuration interaction singles and doubles (CISD). When using MP2, the first step is to run an MP2 calculation for a given geometry in a specified basis and obtain the t2 amplitudes. The remaining steps in the proposed approach follow with the use of these amplitudes. In the case of CISD, the coefficients of the CISD ground state wavefunction (that are directly related to the excitation amplitudes) can be used for finding the dominant excitations, and hence active spaces with good correlation. A second challenge is with respect to the calculation of correlation factor for each possible combination of AS, where the number of such combinations can become prohibitively large when the total number of orbitals ($N$) is large. In such a situation, we suggest the following: we begin by choosing an initial active space AS\textsubscript{I} either based on chemical intuition (such as freezing the core and few lowest energy orbitals) or using perturbation theory-based methods to filter out very weakly correlated orbitals \cite{activespacefinder}. This would reduce the dimensionality of the subspace where possible combinations of orbitals and their correlation factors are computed. The second step would be to run CCSD on the chosen AS\textsubscript{I}, followed by the remaining steps outlined in our method to determine the correlation factors.
 
It is worthwhile to compare our approach with some of the existing strategies for selection of active spaces in the literature \cite{activespacefinder,stein2016automated}. We note a similar but not identical approach to the selection of active orbitals in an earlier work\cite{gao2021computational}. In our paper, we choose the minimal STO-6G basis set for all VQE calculations, which enabled us to perform CCSD calculations smoothly on a classical computer. Other computational details are provided in the next subsection.

\subsection{Computational Details}\label{subsec:Computational Details}
All the electronic structure calculations that were performed on classical computers using methods like full configuration interaction (FCI), CCSD, and complete active space configuration interaction (CASCI) were conducted using the PySCF package \cite{sun2018pyscf}. These results were later used as a benchmark for the quantum computing results. The one- and two-electron integrals, which are required to define the second quantized electronic Hamiltonian, were also computed using the PySCF package in the HF molecular orbital basis. To map the electronic structure problem from the second-quantized form to the qubit space, and to construct the UCCSD and PGSD ansatzes, we used the latest Qiskit 1.1.0 software toolkit \cite{qiskit2024}. While working with UCCSD circuits, we used the default setting of single Trotter step, as implemented in Qiskit.

For all the noiseless VQE simulations, the classical L-BFGS optimizer as implemented in SciPy package was used \cite{virtanen2020scipy}. In these simulations, first, the VQE parameters were initialized to zeroes for the equilibrium geometry. Next, the optimized parameters of the equilibrium geometry were used as the initial parameters for the geometries that are adjacent to the equilibrium geometry. Later, this process is repeated for all other geometries, i.e., the optimized parameters of a geometry are used as the initial parameters for the adjacent geometries. Choosing the initial points in this manner facilitated quicker convergence of the VQE algorithm with both ansatzes, specifically near the bond dissociation regimes. In the noisy VQE simulations, we used the gradient-free COBYLA optimizer (which avoids the errors that arise due to the calculation of gradients for noisy cost functions). In these simulations, the optimized parameters from the noiseless simulations were used as the initial parameters for both UCCSD and PGSD ansatzes. We set the maximum number of iterations of L-BFGS and COBYLA optimizers to 200 and 15000 iterations, respectively, and the maximum absolute value of any component of gradient of energy with respect to the circuit parameters (‘gtol’ argument of L-BFGS optimizer in SciPy) to $10^{-6}$ Hartree/rad for terminating the L-BFGS optimizer.

To assess the quality of the bond dissociation profiles computed using VQE with UCCSD and PGSD ansatzes, we used two metrics: (a) root mean square error (RMSE) and (b) non-parallelity error (NPE). Here, NPE is defined as the difference between the maximum and minimum deviation in the UCCSD/PGSD energies from the CASCI energies (computed over the entire bond dissociation profile). We chose the CASCI results as our benchmark since CASCI is equivalent to performing FCI or exact diagonalization in a chosen active space. We also performed active space CCSD (AS-CCSD) simulations as implemented in PySCF. For compiling the circuits, Qiskit's built-in transpiler is used, and it is configured with all the properties of \textit{\textit{ibm\_brisbane}} device, such as the set of basis gates and coupling map, and its optimization level is set to 3. 

\section{Results and Discussion} \label{sec: Results}
\begin{table*}[h!]
    \centering
    \caption{Quantum resource estimates and the corresponding energy errors of UCCSD and PGSD ansatzes for various ASs. The resource estimates are measured after compiling each of the ansatzes to the basis gate set and coupling map of the \textit{\textit{ibm\_brisbane}} device. The energy errors are obtained while computing the bond dissociation profiles of various molecules using the VQE algorithm on a statevector simulator. Description of the abbreviations - TQG: two-qubit gates, RMSE: root mean square error, NPE: non-parallelity error, mHa: milli Hartree. }
    \resizebox{\textwidth}{!}{%
    \begin{tabular*}{\linewidth}{@{\extracolsep{\fill}}@{}ccccc@{}}
        Molecule & Active Space & & UCCSD & PGSD\\ \toprule
         & & Number of Qubits & 10 & 10 \\ & & Number of Parameters & 54 & 54\\ & & Circuit Depth & 16416 & 5396\\ & & Number of TQGs & 3718 & 1763\\
         H\textsubscript{2}O & AS(6e, 5o) & Total Gates & 27924 & 11837 \\ & & Depth Reduction (\%) & - & 67.13 \\ & & TQG Reduction (\%) & - & 52.58 \\  & & RMSE (mHa) & 1.45 & 2.90\\ & & NPE (mHa) & 4.83 & 8.97 \\ 
         \midrule
         & & Number of Qubits & 8 & 8 \\ & & Number of Parameters & 26 & 26\\ & & Circuit Depth & 5676 & 1800\\ & & Number of TQGs & 1319 & 676\\
         N\textsubscript{2} & AS(4e, 4o) & Total Gates & 10062 & 5165 \\ & & Depth Reduction (\%) & - & 68.29 \\ & & TQG Reduction (\%) & - & 48.75 \\  & & RMSE (mHa) & 2.62 & 2.12\\ & & NPE (mHa) & 5.00 & 5.00 \\ 
         \midrule
         & & Number of Qubits & 12 & 12 \\ & & Number of Parameters & 68 & 68\\ & & Circuit Depth & 24832 & 8027\\ & & Number of TQGs & 5358 & 2329\\
         O\textsubscript{2} & AS(8e, 6o) & Total Gates & 42581 & 17711 \\ & & Depth Reduction (\%) & - & 67.67 \\ & & TQG Reduction (\%) & - & 56.53 \\  & & RMSE (mHa) & 8.66 & 19.62\\ & & NPE (mHa) & 22.35 & 40.30 \\
         \midrule
         & & Number of Qubits & 6 & 6 \\ & & Number of Parameters & 8 & 8\\ & & Circuit Depth & 1286 & 593\\ & & Number of TQGs & 240 & 161\\
         H\textsubscript{2}O & AS(2e, 3o) & Total Gates & 2100 & 1256 \\ & & Depth Reduction (\%) & - & 53.88 \\ & & TQG Reduction (\%) & - & 32.92 \\  & & RMSE (mHa) & $4.015\times 10^{-5}$ & $1.668\times 10^{-5}$\\ & & NPE (mHa) & $1.066\times 10^{-4}$ & $4.554\times 10^{-5}$ \\
         \bottomrule
    \end{tabular*}%
    }
    \label{tab: Quantum Resource Estimates of UCCSD and PGSD}
\end{table*}

In this section, we present and analyze the results obtained by applying the methods described in Sec.~\ref{sec:Methodology} for two scenarios: (a) Noiseless VQE simulations and (b) Noisy VQE simulations. Here, we first performed the noiseless simulations to demonstrate the accuracy of PGSD circuits in estimating the ground-state energies of a given molecule in a chosen AS under ideal (or fault-tolerant) conditions. Next, we conducted the simulations by including various forms of quantum noise to show how PGSD circuits, with their lower quantum resource consumption, can outperform UCCSD circuits and yield lower error rates. To demonstrate the accuracy and efficiency of PGSD circuits, we computed the dissociation profiles of small molecules, in particular, water, molecular nitrogen, and molecular oxygen, choosing different ASs.

\subsection{\label{subsec: Noiseless VQE Simulations}Noiseless VQE Simulations}
\subsubsection{\label{subsec:H2O Dissociation Profile - Noiseless}H\textsubscript{2}O: Symmetric Dissociation Profile}
\begin{figure}[h!]
    \centering
    \includegraphics[width=1.0\columnwidth]{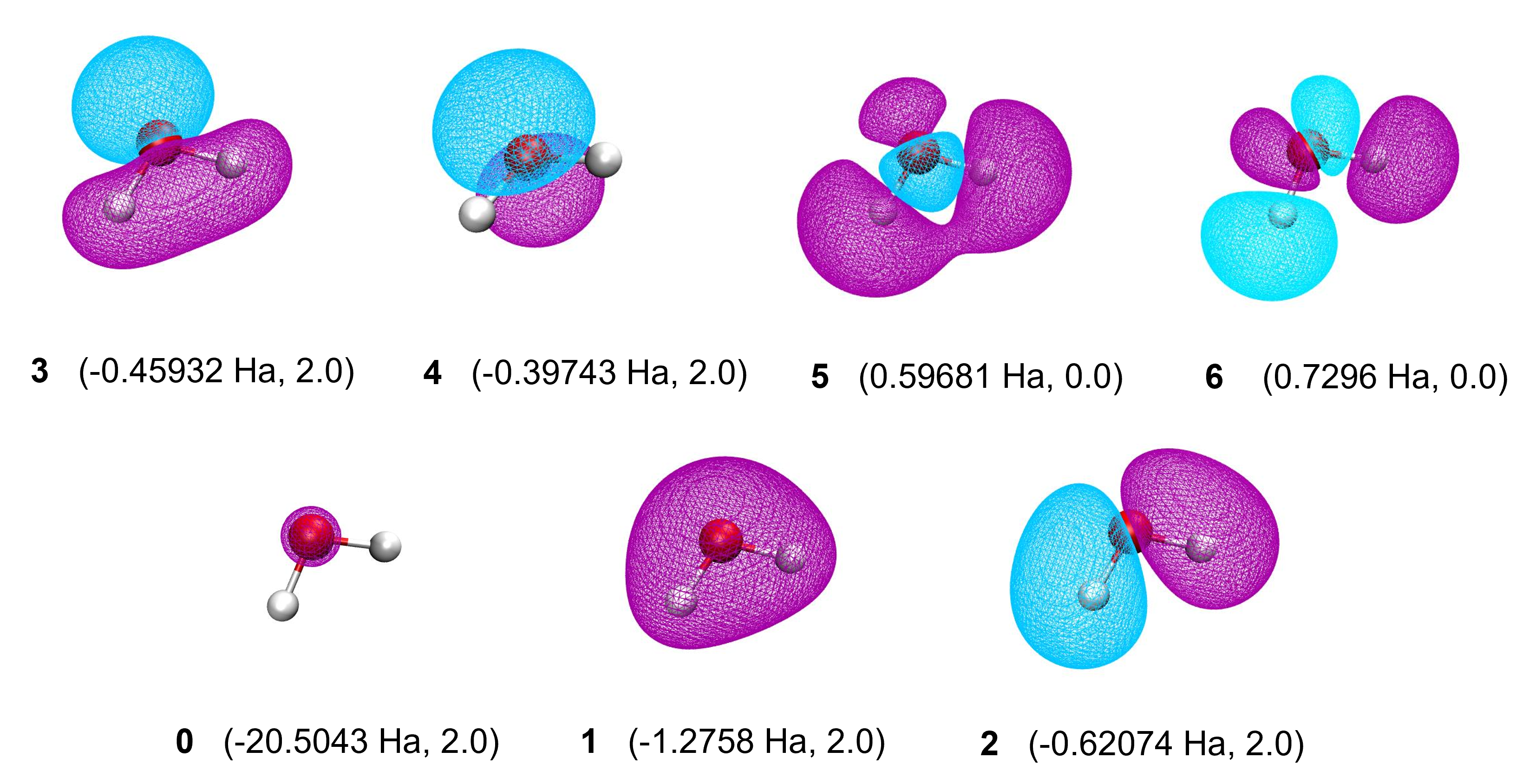}
    \caption{Visualization of the MOs obtained from an RHF calculation of water using STO-6G basis at $R_{OH}=0.958~\angstrom$. The text below each MO indicates the orbital index (starting from zero), the orbital energy, and the occupation number (zero for unoccupied or two for occupied).}
    \label{fig: H2O RHF MOs STO-6G}
\end{figure}

The symmetric dissociation profile of H\textsubscript{2}O, where its ground-state energy is computed at different oxygen-hydrogen distances ($R_{OH}$) for a fixed H-O-H angle, represents a strongly-correlated problem, specifically at large $R_{OH}$ values. In our calculations, the H-O-H bond angle is fixed at $104.478 \degree$ (experimental equilibrium value), and the two hydrogen atoms are symmetrically displaced away from the oxygen atom from $0.5~\angstrom$ till $2.4~\angstrom$. For each $R_{OH}$ distance, a restricted HF (RHF) calculation is performed within the STO-6G basis, which yields seven MOs, and these MOs for the experimental equilibrium bond distance ($R_{OH}=0.958~\angstrom$) are depicted in Fig.~\ref{fig: H2O RHF MOs STO-6G}. 

\begin{figure}[htb]
    \centering
    \includegraphics[width=\columnwidth]{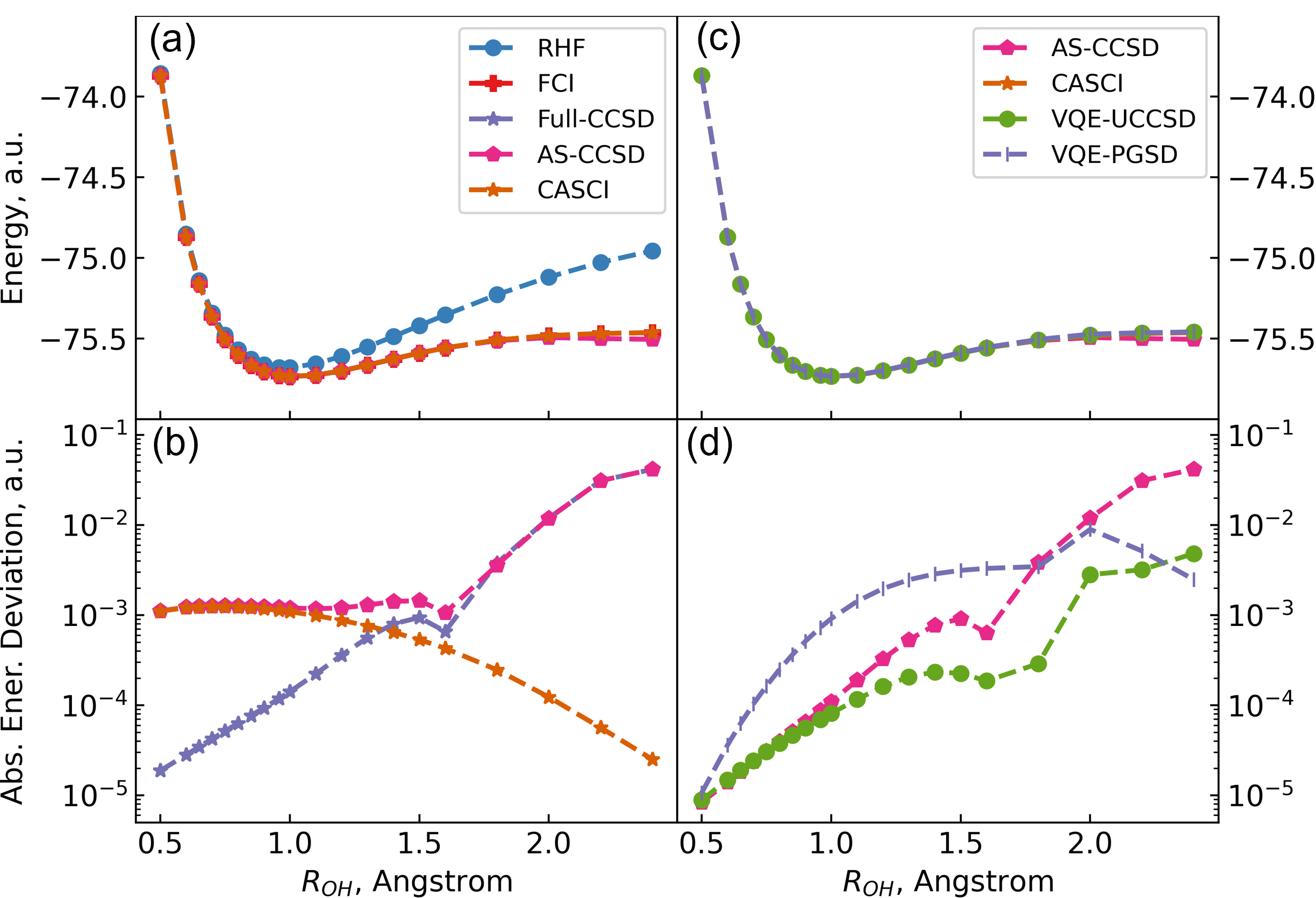}
    \caption{Symmetric dissociation profiles of H\textsubscript{2}O computed with (a) RHF, CCSD, FCI and CASCI methods (b) VQE using UCCSD and PGSD ansatzes at a constant H-O-H angle of $104.48 \degree$. The CASCI, AS-CCSD and VQE calculations are performed in an AS(6e, 5o), where the orbitals are chosen dynamically (as explained in the text). (c) shows the absolute deviation of Full-CCSD, AS-CCSD and CASCI relative to FCI, whereas (d) shows the absolute deviation of VQE energies computed with UCCSD and PGSD ansatzes (using Qiskit) and AS-CCSD energies (using PySCF) relative to CASCI.}
    \label{fig: H2O-STO-6G-Symm-Dissoc-Profile-Quantum}
\end{figure}

The dissociation profiles obtained with various electronic structure methods are shown in Fig.~\ref{fig: H2O-STO-6G-Symm-Dissoc-Profile-Quantum}(a), where we conducted both full space (with seven spatial orbitals and 10 electrons) and active space calculations (with an AS(6e, 5o)). To identify the orbitals that recover the maximum correlation energy for an AS(6e, 5o), we applied the method discussed in Sec.~\ref{subsec:ASS-CCSD} at each bond distance. The different choices of (6e, 5o) ASs and their associated correlation factors and correlation energies calculated at the CCSD level of theory are given in Table \ref{tab: H2O CAS(6e, 5o) AS at equilibrium geometry} for $R_{OH} = 0.958 \angstrom$. For brevity, we only present the data associated with the top 5 choices in Table \ref{tab: H2O CAS(6e, 5o) AS at equilibrium geometry}. The complete data can be found in Table \ref{supp-tabS1: H2O Full CAS(6e, 5o) AS at equilibrium geometry} of the supplementary information.

\begin{table}[h!]
    \centering
    %\small
    \caption{(6e, 5o) ASs of H\textsubscript{2}O at $R_{OH} = 0.958 \angstrom$ in STO-6G basis. The correlation factor $\epsilon$, and correlation energy $E_{corr}$ (in milli-Hartrees, mHa) of each AS computed using the CCSD method are also provided.}
    \begin{tabular}{@{}cccc@{}}
        \toprule
        Active orbitals & $\epsilon$ & $E_{corr}$ (mHa)\\
        \midrule
        (1,2,3,5,6) & 0.0257933 & -48.8990\\
        (2,3,4,5,6) & 0.0201198 & -34.0843\\
        (0,2,3,5,6) & 0.0193250 & -32.9384\\
        (1,2,4,5,6) & 0.0131783 & -29.1106\\
        (0,1,2,5,6) & 0.0123842 & -28.0907\\
        \bottomrule
    \end{tabular}
   \label{tab: H2O CAS(6e, 5o) AS at equilibrium geometry}
\end{table}

From Table \ref{tab: H2O CAS(6e, 5o) AS at equilibrium geometry}, it is evident that the set of orbitals with higher correlation energy ($E_{corr}$) also has a larger correlation factor ($\epsilon$) associated with it, proving that the correlation factor serves as a good metric for identifying the ``orbital set'' that can retrieve significant correlation energy. Using this procedure, we identified the correct set of active orbitals (which provides the highest correlation factor) at every $R_{OH}$ value dynamically. Below $1.4~\angstrom$, we find that the orbital set (1,2,3,5,6) retrieves most of the correlation energy. Similarly, between $1.4 - 1.5~\angstrom$ and above $1.5~\angstrom$, the orbital sets (1,2,4,5,6) and (1,3,4,5,6) retrieve the maximum amount of correlation energy, respectively. It is important to note that choosing the orbital set dynamically (i.e., depending on the geometry) yields smooth energy profiles, unlike the ones obtained by fixing the same orbital set for all distances, which show sudden discontinuities in the energy profiles \cite{rossmannek2021quantum}. As such, our method has an apparent advantage in obtaining accurate potential energy surfaces for a given AS. Since our method provides a systematic approach to identify the correct orbital set that can retrieve maximum correlation for a given AS, it can be extremely useful to benchmark other methods that predict the active set of orbitals based on chemical intuition (for example, using the symmetry of orbitals \cite{eddins2022doubling, motta2023quantum}). 
%Moreover, for the systems studied in this work, since the computational cost associated with performing full CCSD calculations is not enormous, we found that this method is helpful in establishing the effectiveness of PGSD ansatz. 
%The dissociation profiles obtained with conventional electronic structure methods are shown in Fig.~\ref{fig: H2O-STO-6G-Symm-Dissoc-Profile-first}.

From Fig.~\ref{fig: H2O-STO-6G-Symm-Dissoc-Profile-Quantum}(a), it can be noticed that the orbital set that we chose at every bond distance for the (6e, 5o) active space is able to recover most of the correlation energy, leading to a very accurate description of the symmetric O-H bond dissociation. To quantify the accuracy of our results, we plotted the absolute energy deviations of the CASCI, Full-CCSD and AS-CCSD results from the FCI energies in Fig.~\ref{fig: H2O-STO-6G-Symm-Dissoc-Profile-Quantum}(b). Clearly, both Full-CCSD and AS-CCSD results deviated largely (10-100 milli-Hartree) from the FCI energies near the bond dissociation limit. This large deviation is a well-known problem in the literature and it occurs due to the non-variational nature of the single-reference CCSD methods, which causes the energies to diverge toward negative infinity when electron correlation is strong \cite{taube2006new,abrams2005general}. On the other hand, the CASCI energies deviated from the FCI results by a few milli-Hartree at smaller $R_{OH}$ values, and only by a few hundreds of micro-Hartree at larger bond distances. Therefore, when used in conjunction with the CASCI method, the orbital sets that we chose dynamically are able to retrieve the maximum amount of correlation energy. Next, we map these orbital sets to qubits for estimating the ground-state energy of various molecules using the VQE algorithm.

Table \ref{tab: Quantum Resource Estimates of UCCSD and PGSD} summarizes the quantum computational cost (in terms of the number of qubits, parameters, two-qubit gates, circuit depth) of each ansatz and the relative energy errors associated with each AS for different molecules simulated in this work. The gate count and circuit depth estimates for simulating H\textsubscript{2}O in (6e, 5o) active space using both UCCSD and PGSD ansatzes are provided in the first row. Following the procedure outlined in Sec.~\ref{subsec:Givens Ansatz}, the constructed PGSD ansatz consists of 10 qubits, 12 single excitation gates, and 42 double excitation gates, leading to a total of 54 parameters. Although the UCCSD ansatz has equal number of parameters and qubits, its circuit depth and two-qubit gates are higher than the PGSD ansatz by more than $60\%$. Here, we want to emphasize that by using certain qubit mapping techniques (such as parity-mapping), a few qubits can be tapered off from the UCCSD circuit, and accordingly, its circuit complexity can be reduced. However, to make a fair comparison between the PGSD and UCCSD ansatzes, we chose the JW mapping and ensured that the qubit count and number of parameters are the same for both circuits.

Fig.~\ref{fig: H2O-STO-6G-Symm-Dissoc-Profile-Quantum}(c) shows the dissociation profiles of H\textsubscript{2}O computed using the VQE algorithm with both UCCSD and PGSD ansatzes (for an AS(6e, 5o)). Interestingly, unlike the conventional CCSD, the variational character of the UCCSD method \cite{taube2006new} produces better estimates for the ground-state energies near the O-H bond dissociation limit. Moreover, due to the variational nature of the VQE algorithm, both UCCSD and PGSD ansatzes are compelled to provide ground-state energies that are either above or equal to the true ground-state energies. The deviations of UCCSD, PGSD, and conventional CCSD energies from the CASCI energies are plotted in Fig.~\ref{fig: H2O-STO-6G-Symm-Dissoc-Profile-Quantum}(d). At larger bond distances ($R_{OH}>2 \angstrom$), since the third and higher-order electronic excitations are increasingly dominant in the ground-state wavefunction (as can be identified from the CASCI results), the accuracy of UCCSD and PGSD circuits decreases near the bond dissociation limit. However, despite this lack of third and higher-order excitations, both UCCSD and PGSD ansatzes were able to capture more than 500 mHa of correlation energy at the dissociation limit. The amount of correlation energy recovered by both ansatzes as a function of the bond distances are given in Fig.~\ref{supp-fig: Correlation energy of H2O, N2, O2}(a). From these plots, it is clear that the PGSD ansatz is able to very accurately approximate the ground-state wavefunctions almost as good as UCCSD ansatz in the absence of any noise. 

\subsubsection{\label{subsec:N2 and O2 Dissociation-Noiseless}Dissociation profiles of molecular nitrogen and oxygen}
To further investigate the performance of the PGSD ansatz in estimating the ground-state energies for strongly correlated systems, we simulated the double and triple bond dissociation profiles of molecular oxygen (O\textsubscript{2}) and nitrogen (N\textsubscript{2}) in triplet and singlet spin states, respectively. For N\textsubscript{2} in STO-6G basis, the restricted HF calculation yielded 10 molecular orbitals (MOs). Among the 10 MOs, there are two degenerate occupied orbitals and two degenerate virtual orbitals that are oriented perpendicular to the triple bond. Interestingly, towards the dissociation limit, we find that the electronic configurations that arise due to the excitations between these degenerate pairs of orbitals have dominant contributions to the ground-state wavefunction. Moreover, the indices of these MOs change dynamically with the bond distance (since their energies change with distance). Therefore, while constructing the AS, it is crucial to at least include these four orbitals (with correct indices) to capture most of the correlation energy. Accordingly, we chose the (4e, 4o) active space. It is worth highlighting that our method of selecting the active orbitals described in Sec.~\ref{subsec:ASS-CCSD} was able to correctly predict the ASs containing the degenerate orbitals for all bond lengths (for identifying the strongly correlated ASs, we used FCI ground-state wavefunction instead of the CCSD wavefunction, since the later did not converge beyond $1.75\angstrom$). 

\begin{figure}
    \centering
    \includegraphics[width=\columnwidth]{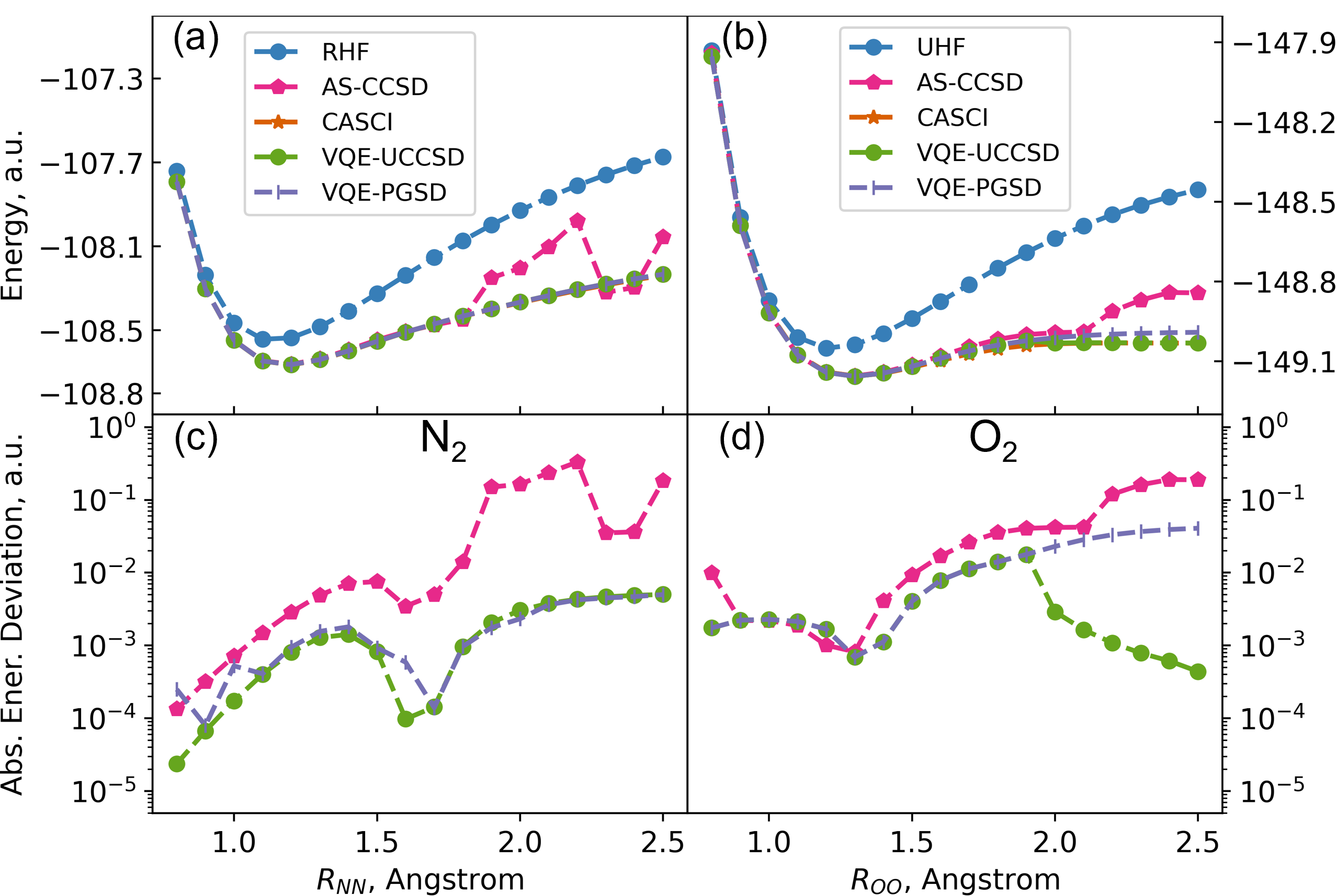}
    \caption{The energy profiles of (a) a singlet N\textsubscript{2} (triple bond stretch) in an AS(4e, 4o) and (b) a triplet O\textsubscript{2} (double bond stretch) in an AS(8e, 6o) computed using various electronic structure methods. The absolute energy errors of UCCSD, PGSD and AS-CCSD methods relative to CASCI for N\textsubscript{2} and O\textsubscript{2} are shown in (c) and (d), respectively. The orbitals in the AS were carefully chosen to include the two pairs of degenerate orbitals (see text for discussion).}
    \label{fig:N2_O2_PES_VQE_UCCSD_PGSD}
\end{figure}

In Figs.~\ref{fig:N2_O2_PES_VQE_UCCSD_PGSD}(a) and \ref{supp-fig: N2 and O2 STO-6G classical methods}(a), we show the energy profiles of N\textsubscript{2} computed using various electronic structure methods by varying its bond length between $0.8 \angstrom$ and $2.5 \angstrom$. Similar to the case of H\textsubscript{2}O, the CCSD energy profile (labeled as Full-CCSD) deviates from the exact (FCI) curve (see Fig.~\ref{supp-fig: N2 and O2 STO-6G classical methods}(a)) in the dissociation limit (above $1.75\angstrom$), and fails to converge for distances beyond $2.1\angstrom$, as also reported in the earlier works.\cite{sokolov2020quantum, rossmannek2021quantum} Similar issues were also observed with the AS-CCSD energy profile (see Fig.~\ref{fig:N2_O2_PES_VQE_UCCSD_PGSD}). The ill-convergence of CCSD and AS-CCSD at these bond distances can be mainly attributed due to their non-variational nature. Moreover, examining the true (FCI) ground-state wavefunctions at these bond distances reveals that these wavefunctions comprise of multiple equally dominating Slater determinants, arising due to the double and higher-order excitations among the degenerate orbitals. In Fig.~\ref{supp-fig: N2 STO-6G FCI ground state wavefunction}, we depicted the dominant electronic configurations and their corresponding coefficients (absolute values) of the FCI wavefunction at $R_{NN}=2.4\angstrom$. The presence of several equally contributing determinants to the ground-state wavefunction indicates the existence of strong static-correlations in the system at these bond distances, making it an ideal system to test the efficacy of PGSD. 
%The presence of several leading determinants in the ground-state wavefunction might necessitate the use of more complex multi-reference (coupled cluster) approaches \cite{} for accurate description of the triple bond breaking/formation in N\textsubscript{2}. 

The energy profiles of N\textsubscript{2} computed for a (4e, 4o) active space using CASCI and AS-CCSD (on a classical computer), and using UCCSD and PGSD circuits (on a state-vector simulator) are shown in Fig.~\ref{fig:N2_O2_PES_VQE_UCCSD_PGSD}(a). The absolute deviations in the energies obtained using AS-CCSD, UCCSD, and PGSD from the standard CASCI method are depicted in Fig.~\ref{fig:N2_O2_PES_VQE_UCCSD_PGSD}(b). As witnessed previously for H\textsubscript{2}O, the obtained PES with UCCSD and PGSD were highly accurate (with only a few milli-hartree deviation from the CASCI across the entire range of bond lengths), indicating their advantage over AS-CCSD (and even Full-CCSD) in simulating systems with static correlation. The quantum computational cost associated with executing UCCSD and PGSD circuits on \textit{\textit{ibm\_brisbane}} quantum device is reported in Table \ref{tab: Quantum Resource Estimates of UCCSD and PGSD}. We find that the depth of PGSD circuit was 68\% lower and approximately 49\% fewer number of two-qubit gates in PGSD circuit for an AS(4e, 4o) compared to UCCSD. Interestingly, with just 30\% of the UCCSD circuit depth, the PGSD circuit was able to estimate energies as good as the UCCSD ansatz (see the NPE and RMSE values).
%When coupled with the VQE algorithm, the variational nature of these ansatzes helps in predicting accurate ground-state energies of molecules.

Next, to demonstrate the competence of PGSD circuits in simulating systems with a triplet spin, we considered the O\textsubscript{2} molecule in its ground-state and computed its dissociation profile. O\textsubscript{2} has 16 electrons, and the unrestricted HF (UHF) calculations in STO-6G basis yields 10 $\alpha$- and 10 $\beta$-spin orbitals, where nine $\alpha$- and seven $\beta$-spin orbitals are occupied. Accordingly, only single excitations are possible in the $\alpha$-spin space, whereas, due to the presence of 3 virtual $\beta$-spin orbitals, up to triple excitations are possible in the $\beta$-spin space. Similar to N\textsubscript{2}, O\textsubscript{2} also has two occupied degenerate orbitals (both in $\alpha$ and $\beta$ spin spaces), and two virtual degenerate orbitals (in the $\beta$ space) that are oriented perpendicular to the bond axis. The computed PESs using UHF, FCI, and CCSD methods (on a classical computer) are shown in Fig.~\ref{supp-fig: N2 and O2 STO-6G classical methods}(b). Once again, CCSD exhibited convergence issues at large bond distances ($R_{OO}\geq 2.0 \angstrom$), leading to significant deviations from the reference FCI calculation. 

Next, for our state-vector simulations, we considered an AS(8e, 6o), including all the degenerate and virtual orbitals. Since among the eight active electrons five are of $\alpha$-spin and remaining are of $\beta$-spin, as part of state initialization in PGSD and UCCSD circuits, we applied Pauli-X gates on the first five spin-$\alpha$ qubits and first three spin-$\beta$ qubits. The single and double excitation gates are applied following the Pauli-X gates to generate the excitations as discussed in Sec.~\ref{subsec:Givens Ansatz}. As before, the resulting PGSD circuit has reduced the circuit depth by 68\% and two-qubit gate count by 56\% compared to the UCCSD version with the same number of parameters (68) and qubits (12) (refer Table \ref{tab: Quantum Resource Estimates of UCCSD and PGSD}). The potential energy curves obtained by employing these ansatzes are shown in Fig.~\ref{fig:N2_O2_PES_VQE_UCCSD_PGSD}(c) together with the CASCI and AS-CCSD calculations in the same active space. Once again, these results reiterate our earlier observations that (i) both UCCSD and PGSD circuits provide more accurate energy estimates than the AS-CCSD, (ii) for strongly correlated systems, due to the presence of the Pauli-Z operators in the fermionic excitation operators accounting for the parity information, UCCSD showed slightly better performance than PGSD toward larger bond distances.

\subsection{\label{subsec: Noisy Simulations}Noisy simulations}
Finally, to compare the performance of the PGSD ansatz over the UCCSD ansatz in the presence of quantum device noise, we performed ground-state calculations using both ansatzes on a noisy simulator. 

Since our aim in performing the noisy simulations is to mimic the behavior of real quantum computers as closely as possible, we configured the Qiskit's \textit{Aer} simulator with various properties, such as the device noise model, basis gate set, and the qubit connectivity (also referred to as coupling map), of the target quantum device (the 127-qubit \textit{\textit{ibm\_brisbane}} machine). Following the workflow of executing quantum circuits on real quantum hardware,\cite{qiskitvqetuto} we transpiled our circuits to match the aforementioned properties of \textit{\textit{ibm\_brisbane}} device, which resulted in 127-qubit circuits with $N_s$ system qubits and $N_a$ ancilla qubits, where the gates are acted only on the system qubits but not on the ancilla qubits. Since simulating such large-qubit circuits on classical hardware requires phenomenal computational power, we removed the $N_a$ ancilla qubits from the 127-qubit circuits without altering any of the system qubits or gates applied to them. In this way, we obtained a reduced $N_s$-qubit circuit compiled to the basis gate set while respecting the coupling map of the target quantum device. We performed all our noisy simulations with the transpiled $N_s$-qubit forms of UCCSD and PGSD circuits (where the value of $N_s$ varies based on the system).

In general, it should be noted that a quantum circuit that has fewer two-qubit gates and reduced depth offers two primary advantages. First, it can be executed more efficiently on a quantum computer, leading to lower computational costs and reduced runtime. Second, and particularly relevant to the NISQ era of quantum computation, it can predict energies and other molecular properties with better accuracy. This is because lower-depth circuits accumulate fewer errors when executed on noisy quantum computers. Together, such circuits improve the affordability and reliability of quantum simulations. Since PGSD circuits have fewer two-qubit gates and reduced depth, they are expected to perform better than the UCCSD circuits in the presence of quantum device noise. 

\begin{figure}[htb]
    \centering
    \includegraphics[width=0.9\linewidth]{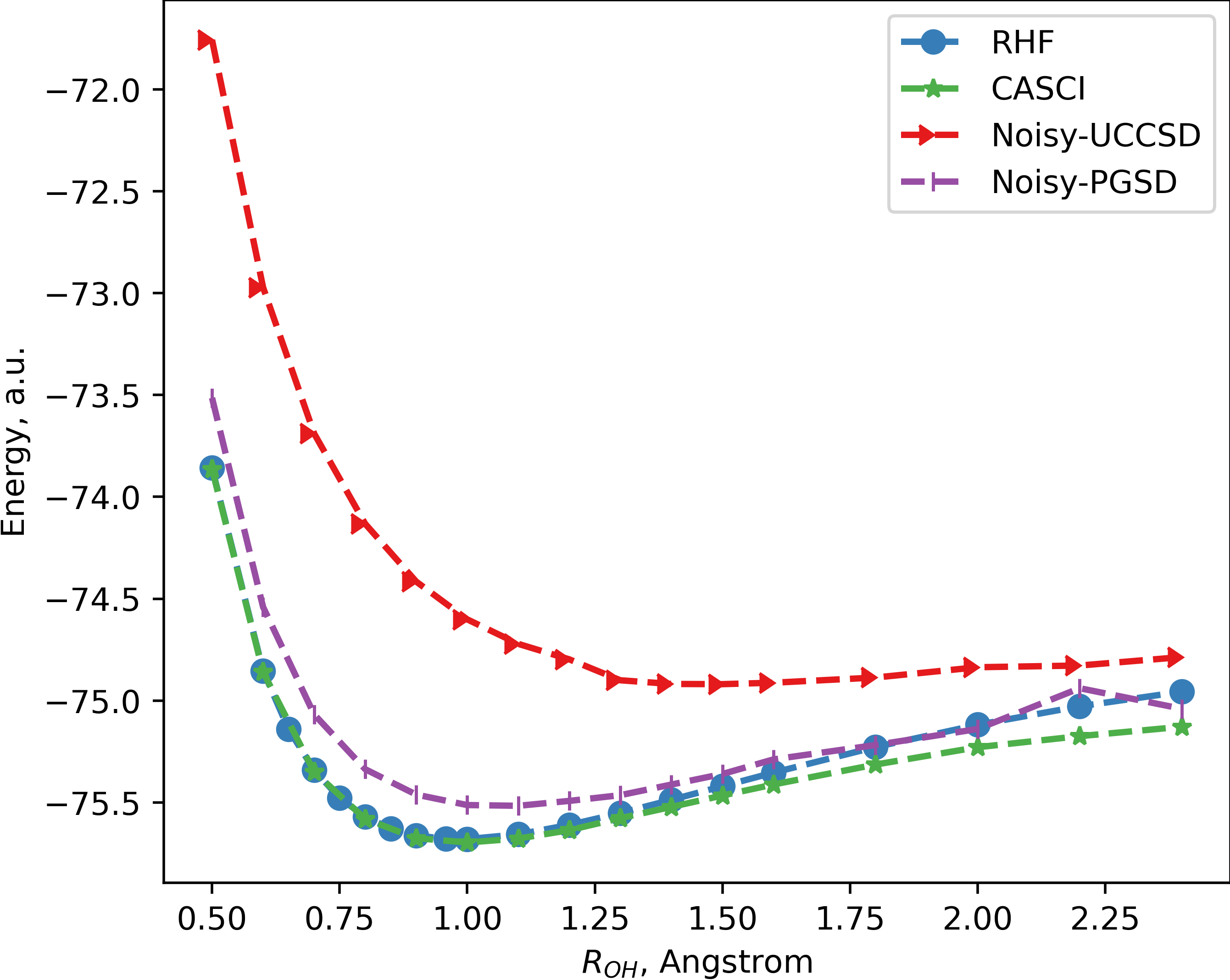}
    \caption{A comparison of PESs of an H\textsubscript{2}O molecule in an AS(2e,3o). The HF and CASCI results were obtained on a classical computer. On the other hand, the Noisy-UCCSD and Noisy-PGSD results were obtained by including the quantum noise of \textit{ibm\_brisbane} device. Quantum simulations were performed using 6-qubit circuits.}
    \label{fig: h2o_noisy_pes_sto6g_cas2e3o_vqe_givens_uccsd}
\end{figure}

To test the above hypothesis, we computed the ground-state energy of an H\textsubscript{2}O molecule in an AS(2e, 3o) with the STO-6G basis set using both PGSD and UCCSD circuits in the presence of quantum noise associated with the \textit{\textit{ibm\_brisbane}} device. During these VQE simulations, we initialized the parameters of each ansatz to their respective optimal values obtained after running VQE on a noiseless simulator. Also, we used 10000 shots to estimate the expectation value of the trial state prepared at each iteration of the VQE algorithm. Under the JW mapping, both PGSD and UCCSD ansatzes produced 6-qubit circuits with properties as reported in Table \ref{tab: Quantum Resource Estimates of UCCSD and PGSD}. Once again, the PGSD circuit exhibited reduced depth (by approx. 54\%) and fewer two-qubit gates (by approx. 33\%) than the UCCSD circuit. The full dissociation profiles of H\textsubscript{2}O simulated with device noise using PGSD and UCCSD circuits together with standard classical methods are shown in Fig.~\ref{fig: h2o_noisy_pes_sto6g_cas2e3o_vqe_givens_uccsd}. In agreement with our hypothesis, compared to the PGSD energy profile, the UCCSD energy profile shows larger deviations from the reference CASCI. 

\begin{figure}[htb]
    \centering
    \includegraphics[width=\linewidth]{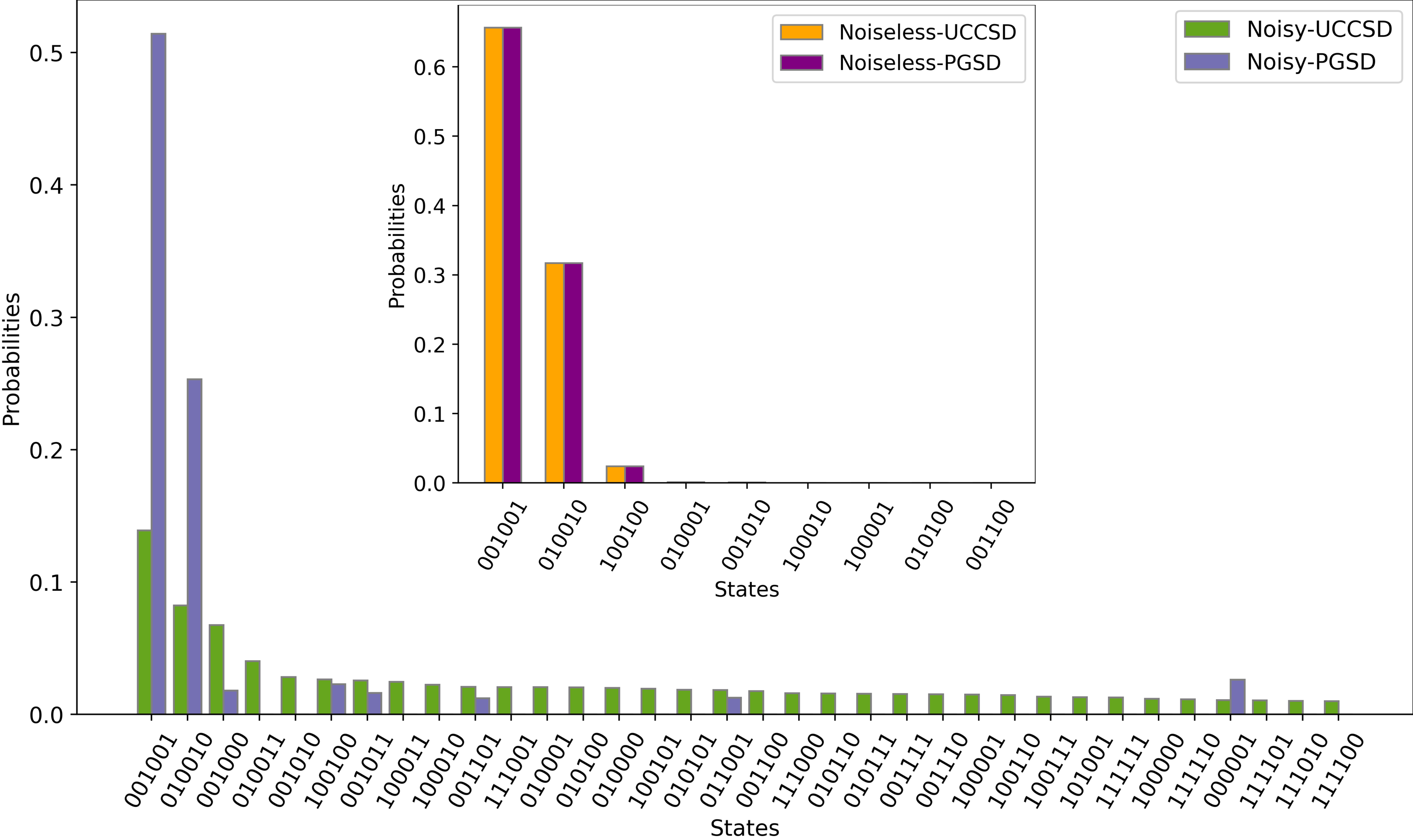}
    \caption{The individual contributions of the Slater determinants (configurations) to the many-body quantum states obtained using the UCCSD and PGSD ansatzes for an H\textsubscript{2}O molecule with $R_{OH}=2.4 \angstrom$ and in an AS(2e,3o). The absolute energy differences between UCCSD and PGSD circuits compared to the CASCI method for this bond length are provided in Table \ref{supp-tab: Noisy-VQE-UCCSD-Givens-CAS(2e, 3o)-H2O}, and discussed in section \ref{supp-sec: Noisy Simulations}. The parent figure (inset) shows the contributions to the quantum state in the presence of \textit{ibm\_brisbane}'s quantum noise (absence of any quantum noise). }
    \label{fig:H2O probability distributions with UCCSD and PGSD}
\end{figure}
To understand the origin of these large energy deviations with the UCCSD circuits, we examined the trial wavefunction prepared by both the ansatzes. In particular, we investigated the effect of circuit depth and gate fidelities on the computation of expectation values using PGSD and UCCSD circuits. For this purpose, we utilized the "Sampler primitive" of Qiskit to sample the quantum states (i.e., to identify the contributions of basis states toward the quantum state) generated by the PGSD and UCCSD circuits (before being acted on by the Hamiltonian). In Fig.~\ref{fig:H2O probability distributions with UCCSD and PGSD}, we present these results for an $R_{OH}$ bond length of 2.4~\AA ~in the presence of device noise. For reference, in the inset of the same figure, we also provided the contributions to the quantum states generated by both ansatzes in the absence of a device noise. Here, the 2.4~\AA ~bond length is considered since the system has high correlation energy at this geometry as shown in Fig.~\ref{supp-fig: Correlation energy of H2O, N2, O2}. 

From the inset of Fig.~\ref{fig:H2O probability distributions with UCCSD and PGSD}, it is clear that at this geometry, the HF state (represented as $001001$) is the most dominant SD in the ground-state wavefunction with 65\% probability, followed by two doubly-excited states (represented by bitstrings $010010$ and $100100$), and negligible contributions from single excitations. However, as shown in the parent figure, where the gate fidelities and qubit coherence times of \textit{\textit{ibm\_brisbane}} device are considered, a large number of qubit basis states, including many unphysical states, contribute to the quantum state generated by UCCSD circuit. Although the HF state is still the dominant SD, its relative probability went down from 65\% to approx. 14\%, and a similar trend is observed with single and double excitations. In contrast, in case of the PGSD circuit, we see fewer number of unphysical states being measured, and more importantly, the relative drop in the probabilities of HF state (as well as other dominant SDs) is much lower compared to UCCSD circuit. Overall, these results indicate that during the state preparation phase in UCCSD, a large number of bit flips occurred due to its higher number of two-qubit gates and larger circuit depth, and these bit flips lead to larger energy errors. On the other hand, PGSD ansatz is able to predict more accurate energy estimates under noisy conditions due to its ability to retain the dominant SDs that are present in the ground-state wavefunction.    

\section{\label{sec: Conclusion} Conclusions}
In this work, we presented methods to (a) identify the correct set of orbitals that can capture the maximum amount of correlation energy for a given size of active space, and (b) construct efficient quantum circuits based on Givens rotation unitaries that can estimate the ground-state energies when combined with the VQE algorithm with high accuracy. Resource estimates for the implementation of this ansatz, termed as PGSD, on a real quantum device show that they possess remarkably lower circuit depths and two-qubit gate counts when compared to the popular UCCSD ansatz. We demonstrated the benefits of the PGSD ansatz by simulating the bond dissociation profiles of water, nitrogen, and oxygen molecules in active spaces of varying sizes both under ideal scenarios and in presence of quantum noise. While the performance of PGSD is comparable to that of UCCSD during the noiseless simulations, it showed a significant improvement over the UCCSD while predicting the ground-state energies of molecules under noisy simulations primarily due to the fewer resource requirement. Also, unlike many hardware-efficient ansatzes, where additional constraints are imposed to preserve various symmetries,\cite{ryabinkin2018constrained, nirmal2024resource} by construction, the PGSD ansatz conserves the particle number and spin symmetries of the trial wavefunction, which makes the classical optimization of its variational parameters faster. Similar to the recently proposed localized unitary cluster Jastrow (LUCJ) ansatz, which was inspired based on the repulsive Hubbard model,\cite{motta2023bridging} the proposed PGSD ansatz strikes a useful balance between the chemically inspired and hardware-efficient ansatzes in terms of accuracy and computational cost.

To further improve the performance of PGSD ansatzes, each single and double excitation operator can be applied more than once. Alternatively, the use of higher-order excitation gates in the PGSD ansatz might enable the PGSD energies to converge to the exact energy as noted previously.\cite{ryabinkin2020iterative} Promising future directions include development of quantum circuits utilizing Givens unitaries for vibrational structure calculations on quantum computers, and we hope that these circuits might reduce the quantum resource estimates compared to state-of-the-art methods.\cite{ollitrault2020hardware} The output state provided by PGSD ansatz can be used as a starting point for calculation of excited states, for e.g., using quantum equation of motion \cite{ollitrault2020quantum}, or for more accurate ground-state energy evaluations with fully quantum approaches, such as quantum phase estimation algorithm.\cite{abrams1999quantum, aspuru2005simulated} Another interesting direction includes, combining the PGSD ansatz with the recently introduced probabilistic configuration recovery scheme,\cite{robledo2025chemistry} where the contributions of the correct dominant configurations to the electronic wavefunction will be recovered, to obtain highly accurate ground-state energies on NISQ devices. 

We showed that our method for selection of active orbitals was able to recover a significant portion of the correlation energy for multiple sizes of active spaces and molecules. After submitting our manuscript, we encountered another useful method for active space selection, namely, the ActiveSpaceFinder (ASF) method,\cite{activespacefinder} which predicts the active space orbitals by first transforming the canonical orbitals to MP2 natural orbitals and then running a Density Matrix Renormalization Group (DMRG) calculation to determine a suitable active space from the set of MP2 orbitals. We conducted a few calculations and found this method to be very useful.
%We do not perform any orbital optimization for the chosen active orbitals as suggested in the literature recently for recovering more correlation \cite{de2023complete}.

\begin{suppinfo}
In the supplementary material, we present the quantum circuit implementations of the single and double excitation gates used in this work as well as the formulas for the number of Slater determinants present in an arbitrary active space. Also provided are the detailed results obtained by applying our method of selection of active spaces and the correlation energy recovered by both UCCSD and PGSD ansatzes to the molecules studied in this work. We report the bond dissociation profiles of N\textsubscript{2} and O\textsubscript{2} molecules computed using various electronic structure methods as well as the results of running VQE with device noise for different sets of initial values for the circuit parameters. 
\end{suppinfo}

\begin{acknowledgement}

S.S.R.K.C.Y. acknowledges the ﬁnancial support of IIT Madras through the MPHASIS faculty fellowship, Centre for Atomistic Modelling and Materials Design, and its new faculty support grants NFSG (IP2021/0972CY/NFSC008973), NFIG (RF2021/0577CY/NFIG008973), and DST-SERB (SRG/2021/001455). We thank Kalpak Ghosh and Sumit Kumar, Department of Chemistry, Indian Institute of Technology Madras, Chennai, India for many helpful discussions.

\end{acknowledgement}

%\input{suppl_info_givens.tex}

%\nocite{*}
\bibliography{achemso-demo.bib}% Produces the bibliography via BibTeX.

\end{document}



\section{Quantum circuit decomposition of the single and double excitation gates used in the PGSD ansatz}\label{sec: Circuit Decomposition}

In this section, we provide the quantum circuit implementations of the single ($U_{SE}$) and double ($U_{DE}$) excitation gates used for the construction of the PGSD ansatz.

\begin{figure}[h!]
    \centering
    \includegraphics[width=0.6\linewidth]{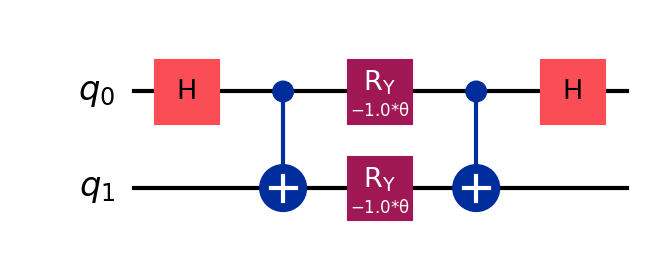}
    \caption{Implementation of the two-qubit single excitation gate.}
    \label{figS1: SE gate}
\end{figure}

\begin{figure*}[htb]
    \centering
    \includegraphics[width=0.9\linewidth]{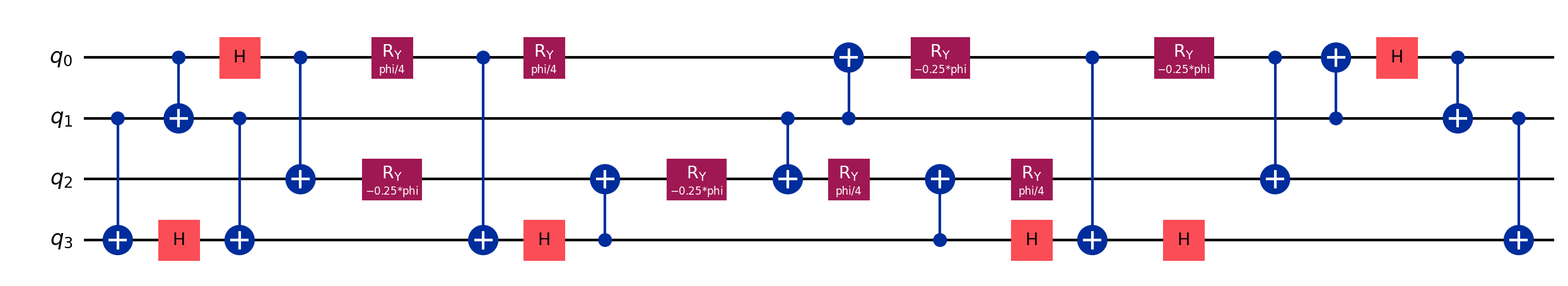}
    \caption{Implementation of double-excitation unitary $U_{DE}$ in terms of single- and two-qubit quantum gates.}
    \label{figS2: DE gate}    
\end{figure*}

\section{Number of Slater determinants present in a given active space}\label{sec: Number of possible excitations}

In this section, we derive the formulas for the number of possible single, double, and triple-order excitations present in a chosen active space comprising of $2N$ electrons and $M$ molecular orbitals, denoted by AS($2N$e, $M$o). Without loss of generality, we fix the number of $\alpha$- and $\beta$-spin electrons to be equal i.e., $N_{\alpha}=N_{\beta}=N$. Since each molecular orbital can be occupied by two electrons of different spins, we have $M$ $\alpha$-spin and $M$ $\beta$-spin orbitals. Out of these $M$ spin orbitals (of any spin), let $M_{occ}$ of them are occupied and $M_{vir}$ are unoccupied (or virtual) orbitals, i.e., $M_{occ}+M_{vir}=M$. Now, for each spin, since we have $N$ electrons that can be filled among $M$ orbitals, the number of possible electronic configurations is equal to $M \choose N$. Similarly, the total number of possible configurations that can be generated with $2N$ electrons and $2M$-spin orbitals is

\begin{equation}
    \centering
    K = {M \choose N}{M \choose N} = {{M_{occ}+M_{vir}} \choose N}{{M_{occ}+M_{vir}} \choose N}
    \label{eqn: K}
\end{equation}

Among the $K$ configurations (or determinants), the number of singly excited determinants (relative to the HF state) is given by
\begin{equation}
    \centering
    S = 2{M_{occ} \choose 1}{M_{vir} \choose 1} = 2M_{occ}M_{vir}
    \label{eqn: S}
\end{equation}
where $M_{occ} \choose 1$ and $M_{vir} \choose 1$ represents the number of ways to remove an electron from one of the $M_{occ}$ orbitals and place it in one of the $M_{vir}$ orbitals, respectively. The factor 2 corresponds to the two possible spin states. Similarly, the number of Slater determinants (SD) that correspond to a double excitation of electrons relative to the HF state is given by 
\begin{align}
    D&=2{M_{occ}\choose 2}{M_{vir}\choose2} + {{M_{occ}\choose 1}^2}{{M_{vir}\choose 1}}^2
    \label{eqn: D1}\\
    &= \frac{1}{2}{[M_{occ}(M_{occ}-1)][M_{vir}(M_{vir}-1)]} + M_{occ}^2 M_{vir}^2 \label{eqn: D2}
\end{align}
where the first term in Eq.~\ref{eqn: D1} represents the excitation of either two spin-up or spin-down electrons, and the second term represents the simultaneous excitation of one spin-up and one spin-down electron. It is important to note that the number of double excitations generated due to the simultaneous excitation of spin-up and spin-down electrons is more compared to those generated by exciting the same spin electrons, which is also evident from Eq.~\ref{eqn: D2}. Similarly, the number of triply excited electronic configurations is given by
\begin{align}
    T =& 2{M_{occ}\choose 3}{M_{vir}\choose 3} \notag\\ 
    &+2{M_{occ}\choose 1}{M_{vir}\choose 1}{M_{occ}\choose 2}{M_{vir}\choose 2}
%      = M_{occ}^2 M_{vir}^2 + M_{occ}M_{vir}(M_{occ}-1)(M_{vir}-1)/2
    \label{eqn: T}
\end{align}
where the first term corresponds to the excitation of either three spin-up or spin-down electrons, and the second term corresponds to the excitation of one spin-up and two spin-down electrons, or vice-versa, together leading to a triple excitation in the overall $2M$-spin orbital space. Evidently, this procedure can be extended to find the number of possible determinants of any higher-order excitation in a given AS.

\section{Actives space selection calculations and results}

In this section, we present the results of applying our method of selecting active spaces to both H\textsubscript{2}O and N\textsubscript{2} molecules  in STO-6G basis for two different sizes of active spaces.

\begin{table}[h!]
    \centering
    %\small
    \caption{(6e, 5o) active spaces of H\textsubscript{2}O at $R_{OH} = 0.958 \angstrom$ in STO-6G basis. The correlation factor $\epsilon$, total ground state energy $E_{tot}$ in Hartrees (Ha), and correlation energy $E_{corr}$ in milli-Hartrees (mHa) computed using CCSD for each active space are also given}
    \begin{tabular}{@{}cccc@{}}
        \toprule
        Active orbitals & $\epsilon$ & $E_{tot}$ (Ha) & $E_{corr}$ (mHa)\\
        \midrule
        (1,2,3,5,6) & 0.0257933 & -75.727687 & -48.8990\\
        (2,3,4,5,6) & 0.0201198 & -75.71287 & -34.0843\\
        (0,2,3,5,6) & 0.0193250 & -75.711727 & -32.9384\\
        (1,2,4,5,6) & 0.0131783 & -75.707899 & -29.1106\\
        (0,1,2,5,6) & 0.0123842 & -75.706879 & -28.0907\\
        (0,2,4,5,6) & 0.0089660 & -75.695507 & -16.7179\\
        (1,3,4,5,6) & 0.008543  & -75.695099 & -16.3104\\
        (0,1,3,5,6) & 0.0077499 & -75.694020 & -15.231\\
        (0,3,4,5,6) & 0.0049414 & -75.6862987 & -7.50987\\
        (0,1,4,5,6) & 0.0021449 & -75.6849409 & -6.15200\\
        \bottomrule
    \end{tabular}
   \label{tabS1: H2O Full CAS(6e, 5o) AS at equilibrium geometry}
\end{table}

\begin{table}[h!]
    \centering
    %\small
    \caption{(4e, 4o) active spaces of N\textsubscript{2} at $R_{NN} = 1.0977 \angstrom$ in STO-6G basis. The correlation factor $\epsilon$, total ground state energy $E_{tot}$ in Hartrees (Ha), and correlation energy $E_{corr}$ in milli-Hartrees (mHa) computed using CCSD for each active space are also given}
    \begin{tabular}{@{}cccc@{}}
        \toprule
        Active orbitals & $\epsilon$ & $E_{tot}$ (Ha) & $E_{corr}$ (mHa)\\
        \midrule
        (4,5,7,8) & 0.0551419 & -108.6438017 & -101.973\\
        (3,5,7,8) & 0.0246365 & -108.588746 & -46.9176\\
        (3,4,7,8) & 0.0246365 & -108.5887463 & -46.9176\\
        (2,5,7,9) & 0.0222668 & -108.592101 & -50.2732\\
        (2,4,8,9) & 0.0222668 & -108.592101 & -50.2732\\
        (5,6,7,9) & 0.0221661 & -108.594995 & -53.1667\\
        (4,6,8,9) & 0.0221661  & -108.594995 & -53.1667\\
        (3,5,7,9) & 0.0220384 & -108.5840027 & -42.174\\
        (3,4,8,9) & 0.0220384 & -108.5840027 & -42.174\\
        \bottomrule
    \end{tabular}
   \label{tab: N2 CAS(4e, 4o) AS at equilibrium geometry}
\end{table}

\section{Correlation Energies recovered by UCCSD and PGSD ansatzes}

We plot the amount of correlation energies recovered by both UCCSD and PGSD ansatzes as a function of bond length.

\begin{figure}[h!]
    \centering
    \includegraphics[width=\linewidth]{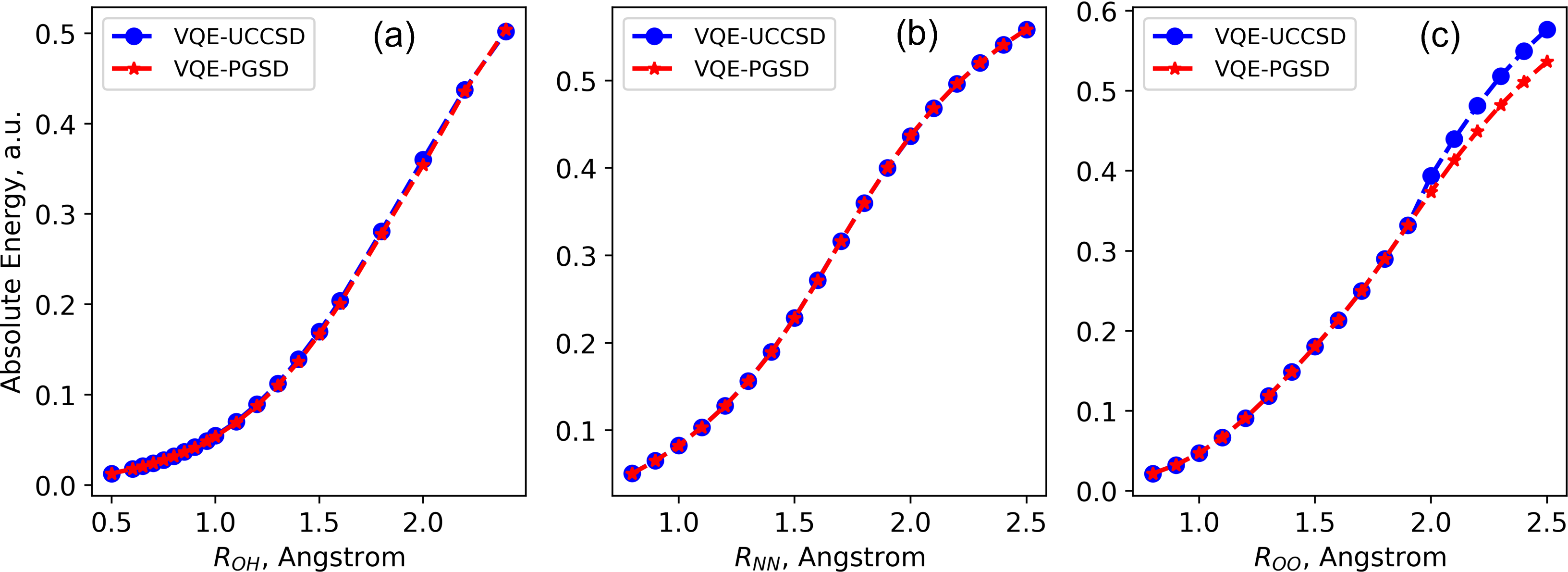}
    \caption{Plot showing the variation of electron correlation energy with bond distances obtained by running VQE with UCCSD and PGSD ansatzes in (a) (6e,5o) active space for H\textsubscript{2}O (b) (4e, 4o) active space for N\textsubscript{2} and (c) (8e,6o) active space for O\textsubscript{2} in STO-6G basis.}
    \label{fig: Correlation energy of H2O, N2, O2}
\end{figure}

\section{Dissociation profiles of molecular nitrogen and oxygen}
In this section, we show the dissociation profiles of N\textsubscript{2} and O\textsubscript{2} molecules computed using various electronic structure methods. We also show the most dominant electronic states and their corresponding absolute amplitudes in the FCI ground state wavefunction of N\textsubscript{2} molecule at $R_{NN}=2.4\angstrom$.

\begin{figure}[h!]
    \centering
    \includegraphics[width=0.6\linewidth]{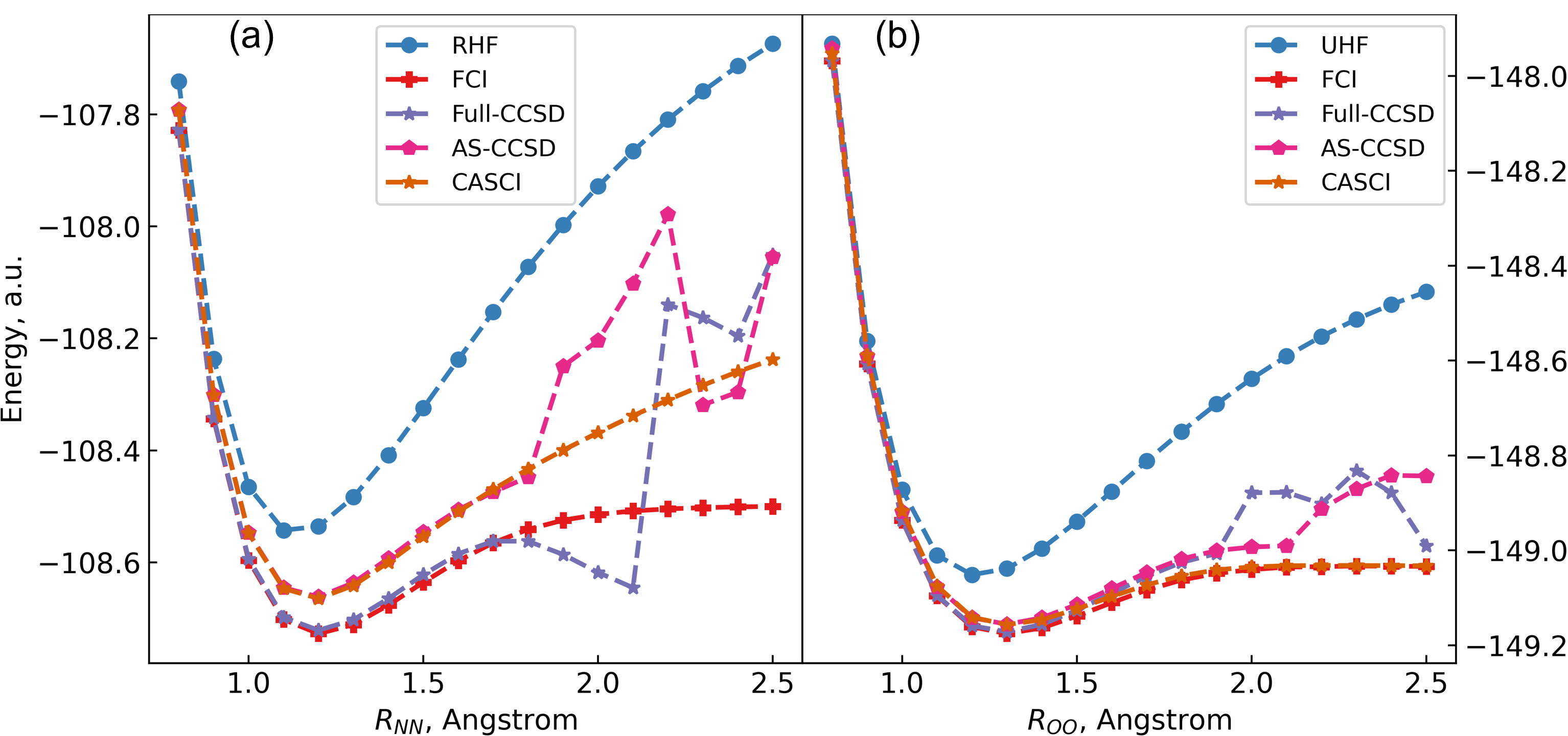}
    \caption{The ground state potential energy curves computed with various electronic structure methods as labeled corresponding to (a) triple bond stretching of N\textsubscript{2} and (b) double bond stretching of O\textsubscript{2} in STO-6G basis. For N\textsubscript{2}, the AS-CCSD and CASCI calculations are performed in a (4e, 4o) active space, while for O\textsubscript{2}, these computations are done considering a (8e, 6o) active space chosen as described in main text.}
    \label{fig: N2 and O2 STO-6G classical methods}
\end{figure}

\begin{figure}[h!]
    \centering
    \includegraphics[width=0.5\linewidth]{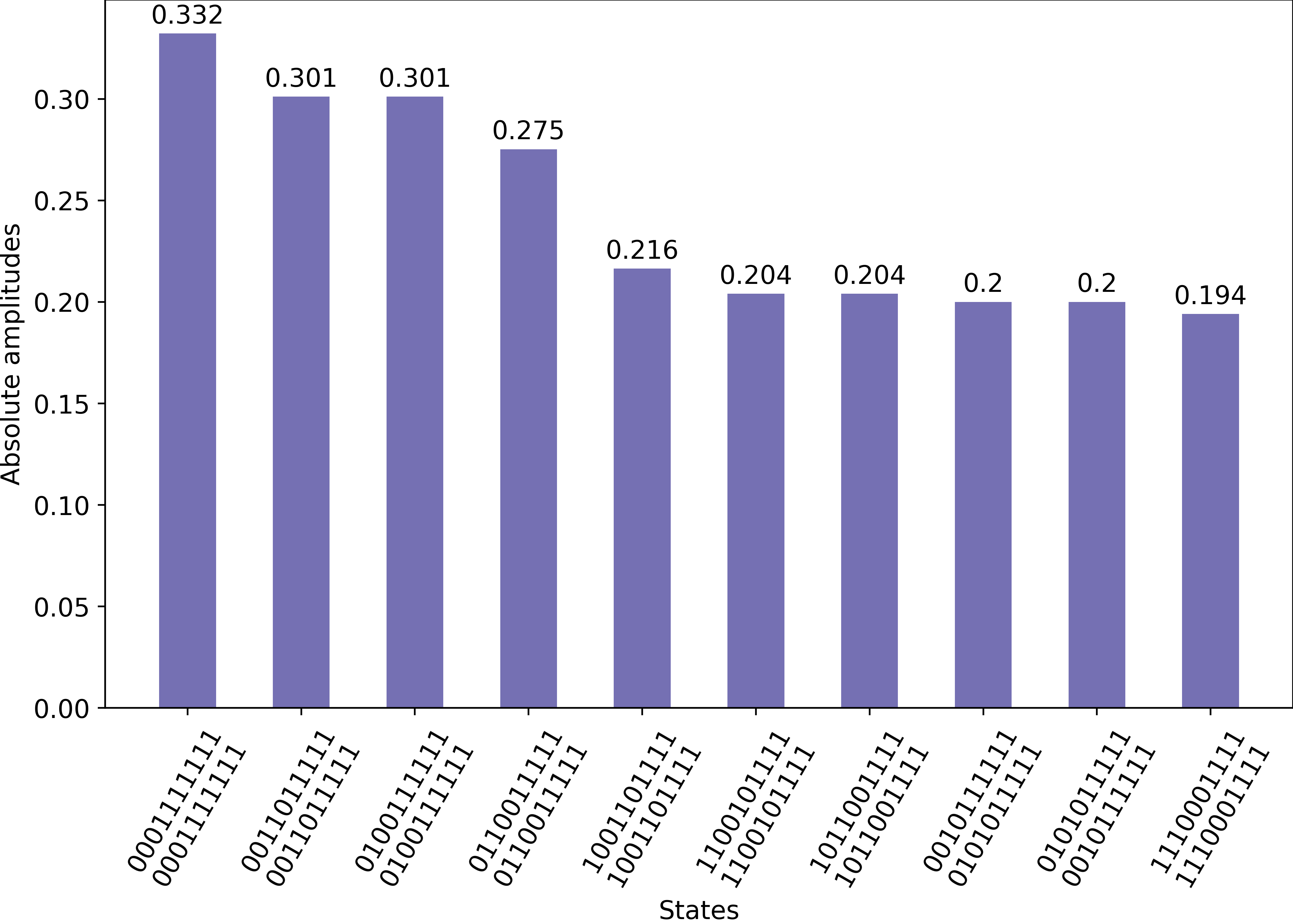}
    \caption{The most dominant electronic states and their corresponding absolute amplitudes in the FCI ground state wavefunction of N\textsubscript{2} molecule at $R_{NN}=2.4\angstrom$. The state with highest amplitude corresponds to the HF state, and the remaining states are a result of electronic excitations among orbitals, primarily between the degenerate set of orbitals.}
    \label{fig: N2 STO-6G FCI ground state wavefunction}
\end{figure}
\newpage

\section{Circuit Parameter Initialization Strategies under Device Noise}
In this section, we present the VQE results computed using different initial parameters for both UCCSD and PGSD circuits. All the simulations were performed by considering the \textit{ibm\_brisbane}'s device noise. 

\begin{figure}[h!]
    \centering
    \includegraphics[width=0.7\linewidth]{figures/H2O_IP_energy_errors_noisy_cas2e3o.png}
    \caption{Absolute energy deviations of VQE energies computed with UCCSD and PGSD ansatzes relative to the CASCI. The circuit parameters are initialized to NOP (Noiseless Optimal Point - obtained from statevector VQE runs), AZ (All zeros), and three different randomly chosen points, denoted as RP1, RP2, and RP3. We considered a (2e,3o) active space for H\textsubscript{2}O and included the device noise of \textit{ibm\_brisbane}.}
    \label{fig: H2O different initialization energy errors under noise}
\end{figure}
\newpage

\section{Noisy Simulations}\label{sec: Noisy Simulations}
Table \ref{tab: Noisy-VQE-UCCSD-Givens-CAS(2e, 3o)-H2O} shows the absolute energy errors obtained with UCCSD and PGSD circuits compared to the CASCI method for the bond length $R_{OH}=2.4\angstrom$. In this table, we presented two different results. First, the energies obtained by running one iteration of the VQE in the noisy environment. These results are shown in the column labeled `Single Iteration'. Second, the energies obtained by running the VQE loop until it converged in the noisy environment. These results are shown in the column labeled `Converged'. In both cases, the UCCSD and PGSD ansatzes were initialized to their respective optimized values obtained from noiseless simulations. 

\begin{table}[h]
    \centering
    \caption{Absolute energy deviations (in milli-hartree) of the computed expectation values using PGSD and UCCSD ansatzes relative to CASCI method for H\textsubscript{2}O molecule in CAS(2e, 3o) active space at $R_{OH}=2.4 \angstrom$ under noisy conditions.}    
    \begin{tabular}{@{}ccc@{}}
        \toprule
        Ansatz & Single Iteration & Converged\\
        \midrule
        UCCSD & 358.591 & 334.350\\
        PGSD & 96.664 & 94.139\\
        \bottomrule
    \end{tabular}
    \label{tab: Noisy-VQE-UCCSD-Givens-CAS(2e, 3o)-H2O}
\end{table}

\begin{figure}[htb!]
    \centering
    \includegraphics[width=0.8\columnwidth]{figures/Fig11_vqe_noisy_opt_prof_uccsd_givens_cobyla_h2o_2.4ang_cas2e3o.png}
    \caption{Plot showing the evolution of energy evaluated as expectation value for the state prepared by PGSD and UCCSD circuits in presence of device noise of \textit{\textit{ibm\_brisbane}} with successive iterations of COBYLA optimizer. This is a result of running VQE for H\textsubscript{2}O at $R_{OH}=2.4\angstrom$ in CAS(2e, 3o) active space. The reference energies (dashed lines) are computed by the standard CCSD and CASCI methods.}
    \label{fig: Noisy_VQE_Opt_profile_Givens_UCCSD_H2O_2.4ang_cas2e3o}
\end{figure}
From the `Single Iteration' results, it is apparent that the ground-state energies predicted by the PGSD ansatz are significantly more accurate compared to the UCCSD ansatz under noisy conditions. To be more specific, the error difference between UCCSD and PGSD predicted energies is as large as 240 mHa! It is crucial to note that the energy obtained from the noisy UCCSD simulations (-74.794157 Ha) is higher than the HF energy (-74.956430 Ha), indicating that conducting UCCSD simulations in noisy environments is counterproductive. On the other hand, the energy obtained from the noisy PGSD simulations (-75.0343675 Ha) could recover about 45\% of the correlation energy (172.076 mHa), making it useful. Interestingly, the `Converged' results are not quite different from the `Single Iteration' results, suggesting that initializing the wavefunction parameters to the noiseless optimal values provides energies close to the converged minimum energy, as observed in some of the earlier works \cite{ghosh2023deep}. As shown in Fig.~\ref{fig: H2O different initialization energy errors under noise}, we also find that this initialization strategy provides the best energies. Fig.~\ref{fig: Noisy_VQE_Opt_profile_Givens_UCCSD_H2O_2.4ang_cas2e3o} shows the VQE optimization profiles with both ansatzes on the noisy simulator. The general observation from Table \ref{tab: Noisy-VQE-UCCSD-Givens-CAS(2e, 3o)-H2O} and Fig.~\ref{fig: Noisy_VQE_Opt_profile_Givens_UCCSD_H2O_2.4ang_cas2e3o} is that device noise deteriorates the accuracy of calculations performed on a quantum computer.

\bibliography{aipsamp}